\documentclass[prl,twocolumn, amsmath,amssymb,superscriptaddress,floatfix]{revtex4-2} %

\usepackage{graphicx}
\usepackage{bm}
\usepackage[dvipsnames]{xcolor}
\usepackage[normalem]{ulem} 
\usepackage{textcomp}
\usepackage{hyperref}
\usepackage{xspace} 
\usepackage{siunitx}
\usepackage{times}
\usepackage{amsmath}
\usepackage{amssymb}
\usepackage{graphicx}
\usepackage{bm}
\usepackage[dvipsnames]{xcolor}
\usepackage[normalem]{ulem}
\usepackage{textcomp}
\usepackage{lineno,hyperref}
\usepackage{hyperref}
\usepackage{xspace}
\usepackage{siunitx}
\usepackage{times}
\usepackage{amsmath}
\usepackage{amssymb}

\newcommand{\WTe}{WTe$_2$\xspace}
\newcommand{\MoS}{MoS$_2$\xspace}
\newcommand{\e}{\'e\xspace}
\newcommand{\EF}{$E_{\rm F}$}

\newcommand{\murm}{%
	\ifmmode
	\mathchoice
	{\hbox{\normalsize\textmu}}
	{\hbox{\normalsize\textmu}}
	{\hbox{\scriptsize\textmu}}
	{\hbox{\tiny\textmu}}%
	\else
	\textmu
	\fi
}

\newcounter{mybibstartvalue}

\begin{document} 
	
	\title{Quantum spin Hall edge states and interlayer coupling in twisted-bilayer 
		\texorpdfstring{\WTe}{WTe2}}
	
	\author{
		\parbox{\linewidth}{\centering Felix L\"upke$^{1, 2, 3, 4}$, Dacen Waters$^{1, 5}$, Anh D. Pham$^{2}$, Jiaqiang Yan$^{6}$, \\ David G. Mandrus,$^{3,6,7}$, Panchapakesan Ganesh$^{2}$,
			Benjamin M. Hunt$^{1, \dag}$\\
			\vspace{5mm}
			\normalsize{$^{1}$Department of Physics, Carnegie Mellon University, Pittsburgh, PA 15213, USA}\\
			\normalsize{$^{2}$Center for Nanophase Materials Sciences, Oak Ridge National Laboratory, \\Oak Ridge, TN 37831, USA}\\
			\normalsize{$^{3}$Department of Materials Science and Engineering, University of Tennessee, Knoxville, TN 37996, USA}\\
			\normalsize{$^{4}$Peter Gr\"unberg Institut (PGI-3), Forschungszentrum J\"ulich, 52425 J\"ulich, Germany}\\
			\normalsize{$^{5}$Department of Physics, University of Washington, Seattle, WA 98195, USA}\\
			\normalsize{$^{6}$Materials Science and Technology Division, Oak Ridge National Laboratory, \\Oak Ridge, TN 37831, USA}\\
			\normalsize{$^{7}${Department of Physics and Astronomy, University of Tennessee, Knoxville, TN 37996, USA}}\\
			{
			}
			\normalsize{$^{\dag}$ E-mail: bmhunt@andrew.cmu.edu}\\
	}}

	\maketitle
	
	\textbf{
		The  quantum  spin  Hall  (QSH)  effect, characterized by topologically protected spin-polarized edge states, was recently demonstrated in monolayers of the transition metal dichalcogenide (TMD) \WTe. 
		However, the robustness of this topological protection remains largely unexplored in van der Waals heterostructures containing one or more layers of a QSH insulator.
		In this work, we use scanning tunneling microscopy and spectroscopy (STM/STS) to explore the topological nature of twisted bilayer (tBL) \WTe which is produced from folded monolayers, as well as, tear-and-stack fabrication.
		At the tBL bilayer edge, we observe the characteristic spectroscopic signature of the QSH edge state 
		that is absent in topologically trivial as-grown bilayer.
		For small twist angles, a rectangular moir\e pattern develops, which results in local modifications of the band structure.
		Using first principles calculations, we quantify the interactions in tBL \WTe and its topological edge states as function of interlayer distance and conclude that it is possible to tune the topology of \WTe bilayers via the twist angle as well as interlayer interactions.
	}
	\begin{figure*}
		\centering
		\includegraphics[width=\textwidth]{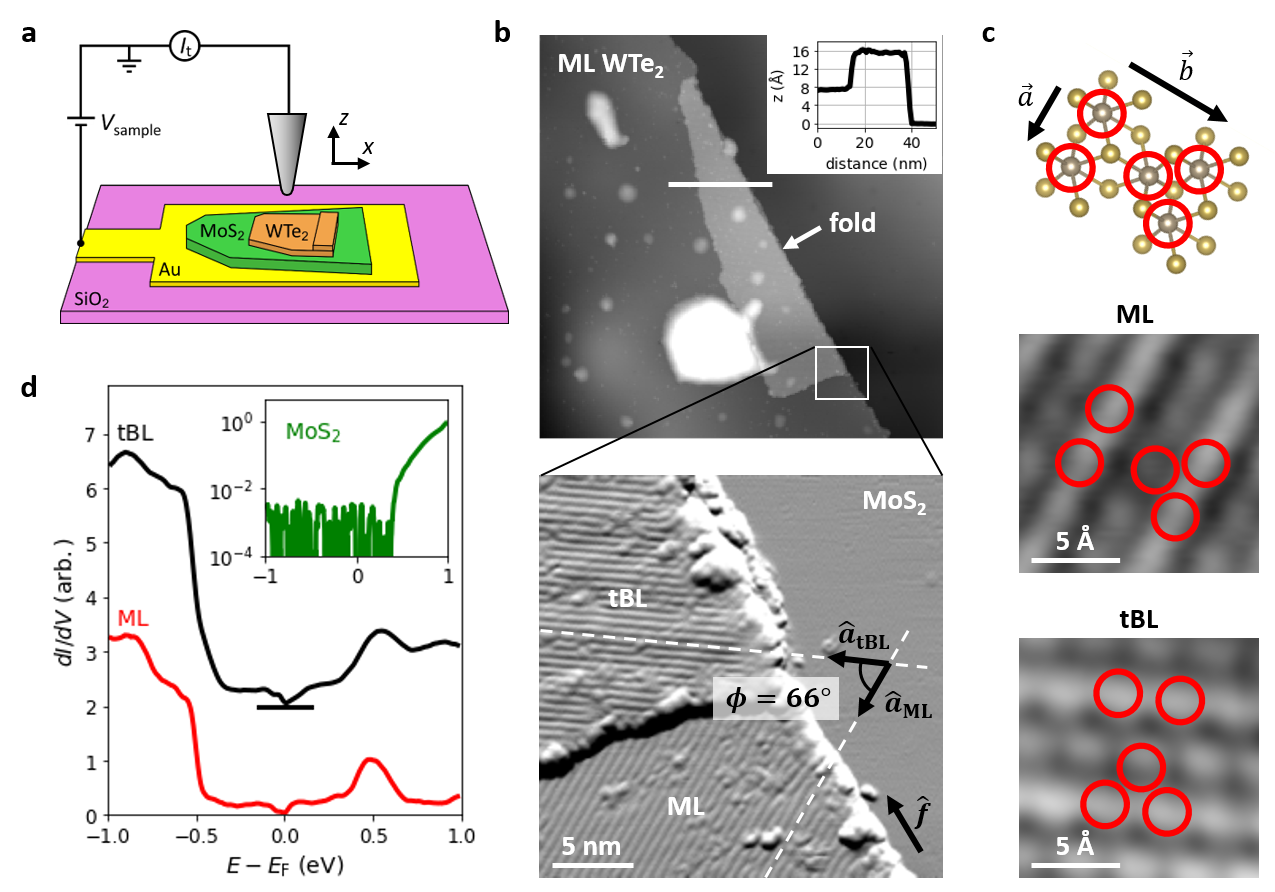}
		\caption{\label{Fig1} \textbf{Twisted bilayer WTe$_2$/MoS$_2$ heterostructure.} \textbf{(a)} Schematic of the tunneling experiment. 
			\textbf{(b)} Large area scan ($200\rm\,nm$) of a \WTe monolayer with a folded edge on a \MoS substrate flake. Inset: Height profile across the folded \WTe bilayer showing layer thicknesses corresponding to single layer step heights of \WTe, indicating atomically clean interfaces. The relative rotational misalignment of the \WTe atomic rows in the two regions is evident in the detail image (gradient image shown), realizing a \WTe twisted bilayer (tBL). {The fold axis $\hat{f}$ is parallel to the resulting \WTe-\MoS step edge}.
			\textbf{(c)} Atomic model of the \WTe unit cell and atomically resolved images from which we determine the lattice misorientation between the two \WTe layers to be $\phi=66^\circ$ (tunneling parameters $V_{\rm sample}=-0.4\rm\,V$, $I_{\rm t}=20$ pA). 
			$|\vec{a}|=\SI{3.48}{\AA}$ and $|\vec{b}|=\SI{6.28}{\AA}$.
			\textbf{(d)} Tunneling spectroscopy on the two \WTe regions reveal almost identical electronic structure, while the \MoS substrate flake shows a band gap around the Fermi energy with the conduction band edge located at $\sim 0.5\rm\,eV$.
		}
	\end{figure*}
	\begin{figure*}
		\centering
		\includegraphics[width=\textwidth]{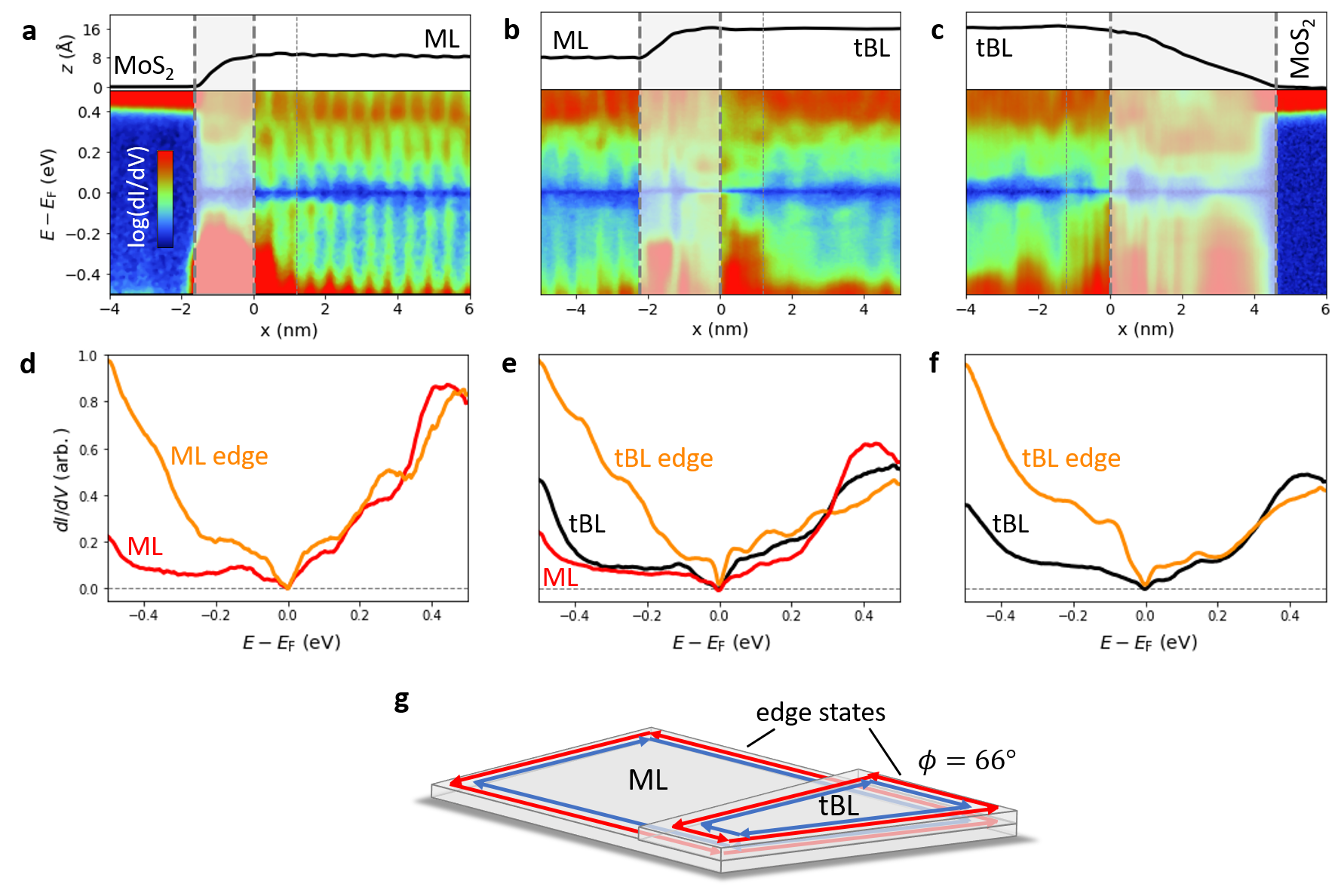}
		\caption{\label{Fig2} \textbf{Observation of edge states in twisted bilayer \WTe.} 
			\textbf{(a)} {Tunneling conductance across the ML \WTe step edge. 
				The location of the topographic edge ($x=0\rm\,nm$) is extracted using a systematic edge detection scheme (see text).
				Right of the edge, the flat terrace of the \WTe monolayer is measured. Here, the QSH edge state consistently appears as an enhanced density of states below the Fermi energy and decays into the monolayer. The decay length is in agreement with the the literature value of $\sim1.2\,\rm nm$ (indicated by a dotted line).
				In the region to the left of the edge, between $-1.5\rm\,nm\lesssim x \leq 0\rm\,nm$, the topography deviates from the flat terrace and the measured signal can be affected by edge reconstructions and artifacts due to the finite size of the STM tip. 
				This `transition region' (overlaid in grey) 
				is therefore excluded from the analysis.}
			\textbf{(b)} Tunneling conductance across the ML-tBL step edge showing the presence of an edge state on top of the tBL terrace. Dashed and dotted lines defined as in (a).
			\textbf{(c)} Tunneling conductance across the tBL-\MoS step edge similarly showing the presence of an edge state on the tBL.
			\textbf{(d-f)} Tunneling spectra extracted from (a-c), averaged over the spatial extent of the indicated edge state regions, and far away from the edges, respectively. The $y$-axis scale is identical for (d-f).
			\textbf{(g)} Schematic of the topological edge states in a rotationally misaligned \WTe homobilayer, reflecting our observations.
		}
	\end{figure*}
	\begin{figure*}
		\centering
		\includegraphics[width=\textwidth]{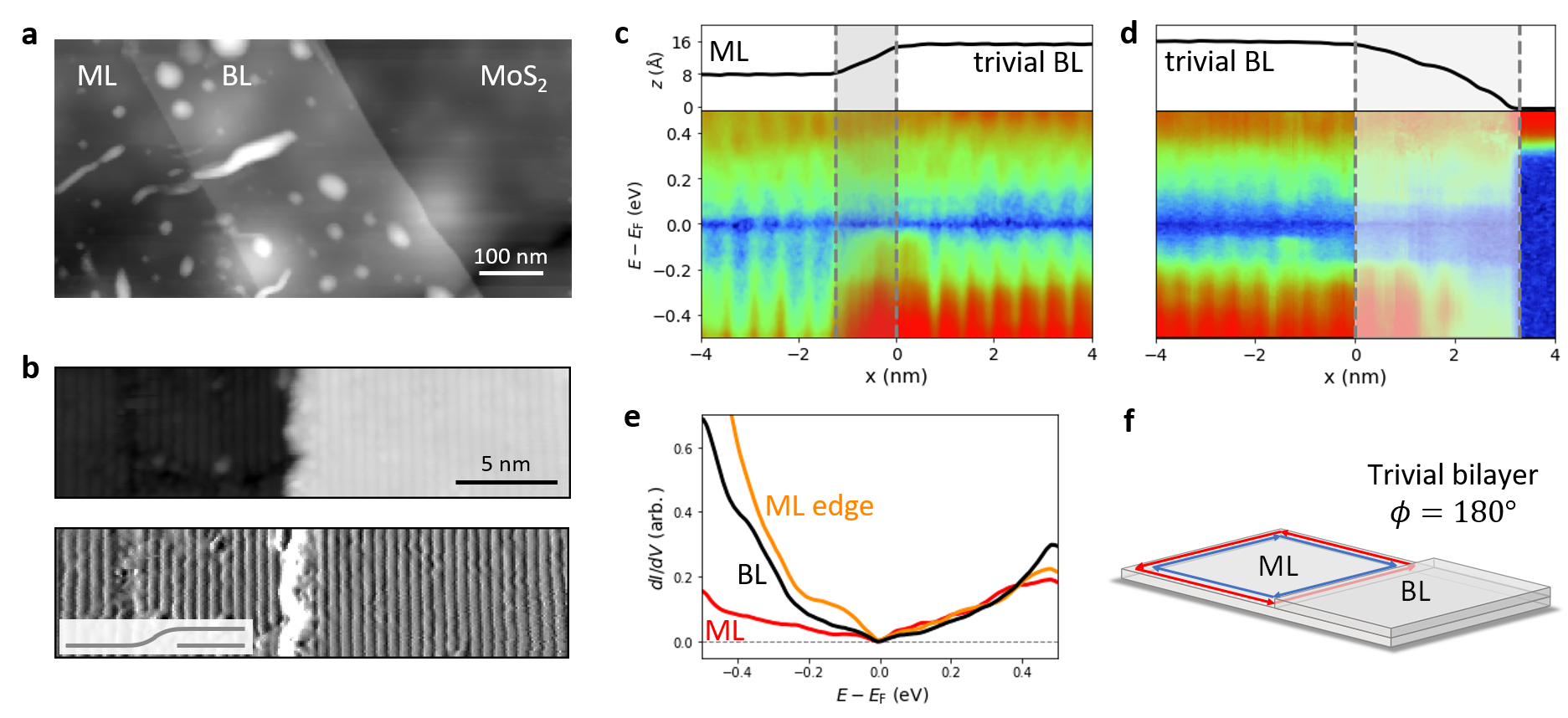}
		\caption{\label{Fig4} \textbf{Edge state at monolayer/as-grown bilayer junction. } \textbf{(a)} Large area scan of an exfoliated as-grown \WTe bilayer connected to the monolayer.
			\textbf{(b)} Atomic resolution scan and gradient of the ML-BL junction showing parallel atomic rows.
			The top layer is expected to be continuous as a result of the sample fabrication technique (inset: schematic of the top layer draped over the second layer underneath).
			\textbf{(c, d, e)} Tunneling conductance measured across the ML-BL edge (c) and BL-\MoS edge (d) showing no signature of an edge state on the BL terrace.
			However, we observe an increase in the density of states in the transition region between the ML and BL which is identical to the signature of the QSH state shown in Fig. 2a.
			The tunneling spectra in (e) are extracted from (c) and (d), averaged over the spatial extent of the indicated edge state region, and far away from the edges, respectively.
			\textbf{(f)} Schematic of the QSH edge state in a \WTe ML-BL junction.
		}
	\end{figure*}


	The variety of layered two-dimensional (2D) materials available for isolation and fabrication into van der Waals heterostructures provides almost endless combinations for device engineering \cite{Geim2013}.  
	1T'-TMDs have attracted recent attention among the layered materials following the prediction of the QSH edge state in the monolayer (ML) limit \cite{Qian2014}. 
	In particular, \WTe has been of great interest due to its realization of multiple exotic physical phenomena depending on its thickness. 
	Monolayer \WTe has been reported to exhibit the predicted QSH effect, which persists up to $100\,\rm K$ \cite{Fei2017, Tang2017, Wu2018, Shi2019}. \WTe monolayers further host a superconducting phase below $\sim 1$ K when electrostatically gated into the conduction band \cite{Sajadi2018, Fatemi2018}. Bilayer (BL) \WTe on the other hand is predicted to be a topologically trivial semi-metal \cite{Muechler2016} which has also been shown to exhibit ferroelectric switching when 
	a surface normal electric field is applied \cite{Fei2018}. In the bulk limit, \WTe was reported to be a type-II Weyl semi-metal and higher order topological insulator with 1D hinge states \cite{Li2017, Choi20}. The high degree of layer tunability in \WTe makes it an attractive candidate for integration in van der Waals heterostructures, e.g. to realize one-dimensional topological superconductivity \cite{Lupke2019}.
	However, there are open questions, such as the effect of coupling to neighboring layers on the topological properties, which have to be thoroughly understood for applications of the topological edge states in electronic devices.
	
	Recently, {the interactions between} rotationally misaligned layers of 2D materials have attracted attention as hosts of a variety of interesting highly-correlated phenomena such as insulating states, superconductivity, and unique topological phases \cite{Cao2018_CI, Cao2018_SC, wang2019magic, Yankowitz1059, Lu2019, Wu2019}. While most studies of twisted bilayers seek to achieve highly-correlated states through a moir\e pattern in which flat bands emerge in the electronic structure \cite{Cao2018_CI, Cao2018_SC, Pan18, Yankowitz1059, Wu2019, Serlin2019, waters2020flatbands}, here we focus on the effects of interlayer coupling on the already present QSH edge state in ML \WTe. We find that for incommensurately stacked twisted bilayer \WTe, both layers retain their topology, which results in two sets of QSH edge states sitting on top of each other.
	{We experimentally study the edge states and interlayer coupling for multiple twist angles and rationalize our results based on first-principles calculations,}
	demonstrating the robust topological protection of the QSH edge state predicted in the literature \cite{Kane2005}. 
	In contrast to twisted bilayers, we do not observe {an edge} state feature at the topologically trivial as-grown bilayer \WTe edge.
	The topologically trivial nature of bilayer \WTe is supported by our observation of a QSH edge state in monolayer \WTe at a monolayer-bilayer junction. 
	
	To explore the QSH edge state in exfoliated \WTe, we study \WTe/\MoS heterostructure{s} (Fig. \ref{Fig1}a). Samples are fabricated using a recently developed dry-transfer flip technique that allows the stacking of van der Waals materials while maintaining atomically clean surfaces and interfaces (for details see Materials and Methods, section S1 and Ref.~\citenum{Lupke2019}). 
	\MoS is used as a substrate flake to provide an atomically flat support for the \WTe. The typical signature of the \WTe QSH edge state in the energy range ($-0.5\,\mathrm{V} \lesssim V_{\mathrm{sample}} \lesssim 0\,\mathrm{V}$) is located in the \MoS band gap, so that substrate effects can be excluded when analyzing the edge state spectra.
	In addition, we can tunnel into the \MoS conduction band at a sample bias of  $V_{\mathrm{sample}}\gtrsim 0.5\,\mathrm{V}$, allowing us to safely scan over the edge of the \WTe flakes.
	
	In large area scans of the heterostructures, we find random folds along the edge of the \WTe monolayer (Fig. \ref{Fig1}b). Such folds occur by coincidence during the exfoliation and dry-transfer and are usually undesired, e.g. for fabrication of transport devices. 
	{However, the fold offers a unique opportunity to study the interaction between the QSH edge states, which are naturally located right on top of one another along the fold.}
	A profile across the folded region shows a ML step height of $\sim 7.5\,$\AA, i.e. slightly larger than for \WTe monolayers on graphene or NbSe$_2$ \cite{Tang2017,Lupke2019} which we attribute to the electronic contrast between the respective substrate materials and the \WTe.
	The step height from the first to the second WTe$_2$ layers is also $\sim 7.5\,$\AA, which slightly deviates from the as-grown bilayer step height ($\sim 7.1\,$\AA\ \cite{Jia2017, Lupke2019}). This deviation can be explained by the incommensurate stacking of the rotationally misaligned layers, resulting in a larger interlayer separation.
	We note that the folding of the \WTe monolayer, due to its resemblance of a screw transformation, in combination with the monolayer 1T' crystal  screw symmetry, $\bar{C}_{2a}=t(\vec{a}/2)C_{2a}$ \cite{Muechler2016}, results in a bilayer that is identical to two monolayers stacked on top of each other with a twist angle $\phi$, {which} is determined by the orientation of the fold with respect to the lattice vectors (see section S2).
	{Using atomic resolution images of the ML and folded region (Fig. \ref{Fig1}c), we determine the relative twist angle between the two layers to be $\phi=|\cos^{-1}\left(\hat{a}_{\rm ML}\cdot\hat{a}_{\rm tBL}\right)|=66^{\circ}=|2\cos^{-1}(\hat{ f}\cdot\hat{a}_{\rm ML})|$, where $\hat{f}$ is the fold axis orientation}. 
	For reference, as-grown bilayer \WTe corresponds to $\phi=180^{\circ}$ by this definition.
	As a result of the large twist angle, we do not observe a pronounced moir\e pattern {and spectra} taken on the folded region are almost identical to that of monolayer \WTe, indicating decoupled layers (Fig. \ref{Fig1}d).
	This observation implies that there should be two independent sets of QSH edge states, one in each layer, which we explore in the following.
	
	First, we examine the ML \WTe edge (Fig. \ref{Fig2}a), where we find the typical spectroscopic signature of the QSH edge state, in the form of an enhanced density of states below \EF, in agreement with the literature \cite{Jia2017, Tang2017, Lupke2019}. 
	The QSH edge state is observed on the flat terrace next to the onset of the topographic step edge, which distinguishes it from edge effects, such as edge reconstructions, and tip artifacts that can occur in the transition region
	due to the finite tip size.
	{
		We systematically identify the positions of the topographic edges -- namely, $x=0\rm\,nm$ and the extension of the transition regions in Figs.~\ref{Fig2}a,b,c and \ref{Fig4}c,d -- as the points where the topography gradient reaches a threshold value of $dz/dx=\pm1.5\,\rm\AA/nm$ (see section S6 for details).
		Using this detection scheme, we find that the spatial extent of the QSH edge state on the ML \WTe terrace is $\sim1.2\rm\,nm$, in agreement with that observed in \WTe on graphene \cite{Tang2017} and on NbSe$_2$ \cite{Lupke2019} and similar to what has been predicted theoretically \cite{Arora20}.}
	
	Turning now to the  ML-tBL edge, we observe a similarly enhanced density of states, which again extends $\sim1.2\,\rm nm$ from the step edge (Fig. \ref{Fig2}b),
	{indicating the presence of an edge state in the upper \WTe layer, since STM is sensitive almost exclusively to the topmost layer.}
	Furthermore, we observe the same increase in density of states at the edge {along} the fold, where the tBL steps down to the \MoS (Fig. \ref{Fig2}c), again with the same lateral extension.
	Along this edge the two sets of QSH edge states sit on top of each other, which is in contrast to interior edges for which the QSH edge states can only interact with the (gapped) underlying bulk.
	Comparing individual spectra extracted from the conductance maps (Fig. \ref{Fig2}d-f), it is evident that the spectra far away from the edge of the ML \WTe and the tBL \WTe are very similar over this energy range. In addition, the spectra at the edges of the ML and tBL \WTe all show a very similar enhanced density of states below \EF, which is consistent with the typical signature of the QSH edge state. 
	We conclude that the individual layers are decoupled, each with {their} own QSH edge state (Fig. \ref{Fig2}g).
	As a result of the weak coupling, we expect that inter-layer scattering vanishes.
	Furthermore, due to the monolayer screw symmetry, the orientation of the spin-momentum locking is identical in both layers in the absence of symmetry breaking fields \cite{Arora20,Zhao21}. In combination with the prohibited intra-layer backscattering, this {suggests}
	strong protection of the QSH edge states at the folded tBL \WTe edge (see also Fig. S6).

\begin{figure}[!h]
		\includegraphics[width=3.5in]{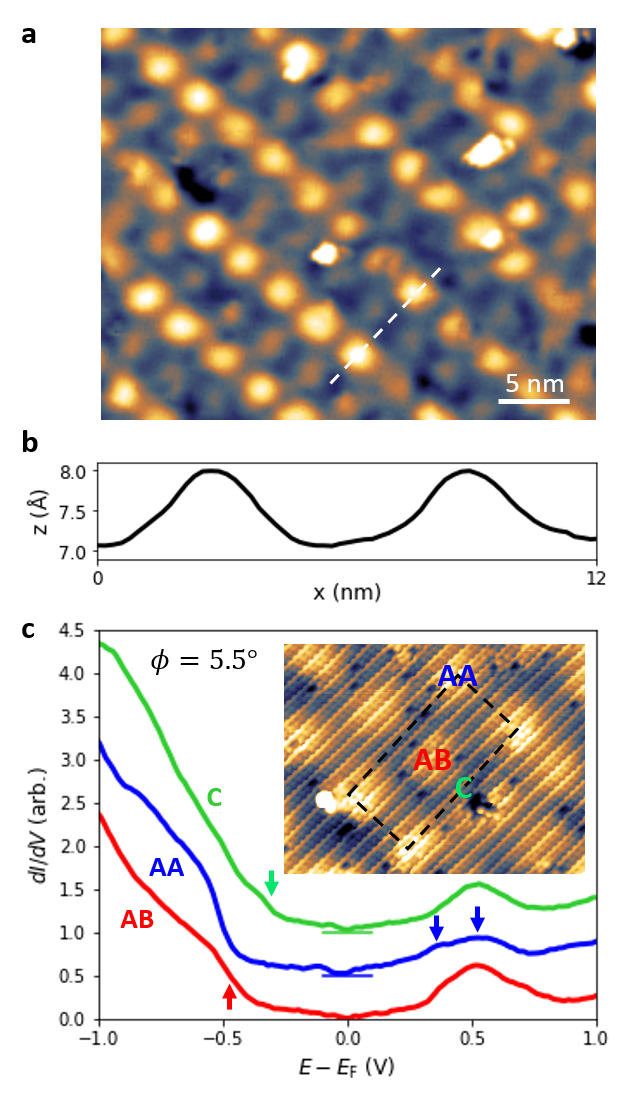}
		\caption{\label{Fig5} \textbf{Tear-and-stack twisted bilayer \WTe.} 
			\textbf{(a)} Large scale topography of the moir\e pattern with periodicity corresponding to $\phi=5.5^{\circ}$ ($V_{\rm sample}=-2\rm\,V$, $I_{\rm t}=20$ pA).	
			\textbf{(b)} Height profile of the moir\e pattern along the dashed line in (a).
			\textbf{(c)} Tunneling conductance of the tear-and-stack tBL \WTe, in AA and AB-stacked regions of the moir\e unit cell. 
			While on the AA site, the conduction band peak is split (blue arrows), on the AB site, the valence band edge is broadened (red arrow), indicating distinct electronic properties within the moir\e unit cell.
			The corresponding spectrum shows an increase in the valence band intensity at $E-E_{\rm F}=-0.3\rm\,eV$ (green arrow) while the conduction band is mostly unchanged. This observation is in agreement with the as-grown \WTe spectra shown in Fig. 4.
			Inset: Atomic resolution topography with the moir\e unit cell indicated in white ($V_{\rm sample}=-1\rm\,V$, $I_{\rm t}=55$ pA).
		}
	\end{figure}

	{Next}, we explore a step edge between a \WTe monolayer and an as-grown topologically trivial bilayer (Fig. \ref{Fig4}a, b). 
	In the present sample, the top \WTe layer of the as-grown bilayer is continuous and draped over the second layer underneath due to the flipping of the sample during fabrication. 
	With multiple iterations of very carefully prepared sharp tips, we are able to resolve the draped monolayer free from significant tip artifacts and measure the tunneling conductance across the ML-BL junction (Fig. \ref{Fig4}c and section S7).
	We find that there is a clear increase in the density of states in the \WTe monolayer just before the bilayer, again with  an extension of $\sim1.2\,\rm nm$.
	The shape of this feature (Fig. \ref{Fig4}c) shows remarkable agreement to that observed at the monolayer \WTe/\MoS edge in Fig. \ref{Fig2}d.
	We therefore conclude that the spectroscopic feature at the ML-BL junction is the manifestation of the QSH edge state between the topological \WTe monolayer and the topologically trivial \WTe bilayer (Fig. \ref{Fig4}f).
	{This finding verifies that there is a significant interlayer interaction in aligned, as-grown bilayer \WTe, which is in contrast to the above studied twisted bilayers. Such an interlayer-interaction-induced change in topology 
		between the monolayer and bilayer is in agreement with \WTe undergoing another topological phase transition at higher layer number, possibly to a higher-order topological insulator \cite{Choi20}.
		The corresponding edge states of bulk \WTe have an electronic structure \cite{Peng2017} that is distinctly different from the edge states observed in ML and tBL \WTe, indicating that these are in fact different topological states.
		
		In order to analyze the effect of inter-layer interactions in greater detail, we further study a tBL sample which we fabricate using a modified tear-and-stack technique (for details see Materials and Methods and section S8).
		{The tear-and-stack sample fabrication allows us to directly control the twist angle, and therefore manifest a moir\'e pattern in which the interlayer coupling is expected to be modulated across the moir\'e unit cell \cite{waters2020flatbands}. }
		The resulting long-range moir\e pattern has a rectangular unit cell and periodicity of $\sim6.5$ nm
		(Fig. \ref{Fig5}a, b). The corresponding twist angle is $\phi\approx5.5^{\circ}$ following the same angle definition as with the folded bilayer, which we also directly confirm at the tBL edge (Fig. S11).
		{Resulting from the different stacking of W atoms, we observe two different corrugation maxima in the moir\e unit cell, {where the A-site W atom in one layer vertically aligns with the A- or B-site W atom of the other layer}, i.e. AA- or AB-stacking, respectively.}
		{Due to the rectangular shape of the moir\e unit cell, the AA-stacked sites form a distinct quasi-one-dimensional pattern, which has been proposed to result in an array of 1D Luttinger liquids at low energy \cite{Wang2021}.}

		Tunneling spectra recorded in different positions of the moir\e unit cell (Fig. \ref{Fig5}c) show distinctly different spectral features. 
		In AA-stacked positions, the conduction band peak at $0.5\rm\,eV$  shows a significant splitting, while spectra taken in AB sites show a stronger hybridization in the valence bands when compared to ML \WTe.
		Spectra taken at C stacked locations in the moir\e unit cell show similarity to that of the as-grown bilayer, with a valence band feature at $\sim-0.3\rm\,eV$.
		Interestingly, this observation {roughly }coincides with the arrangement of W atoms in this location of the tBL moir\e unit cell being similar to that of the as-grown bilayer, despite the C-stacked \WTe having $\sim0^{\circ}$ twist compared to the as-grown bilayer's 180$^{\circ}$ (see section S8 and Fig. S12).
		Since this position also is the minimum of the moir\e corrugation, we expect the hybridization of the two layers to be the strongest throughout the moir\e unit cell.
		In combination, we attribute the distinct changes of the local electronic structure to the different interlayer interactions, which are the result of the local stacking geometry.
	}
	
	\begin{figure}[!h]
		\includegraphics[width=3.4in]{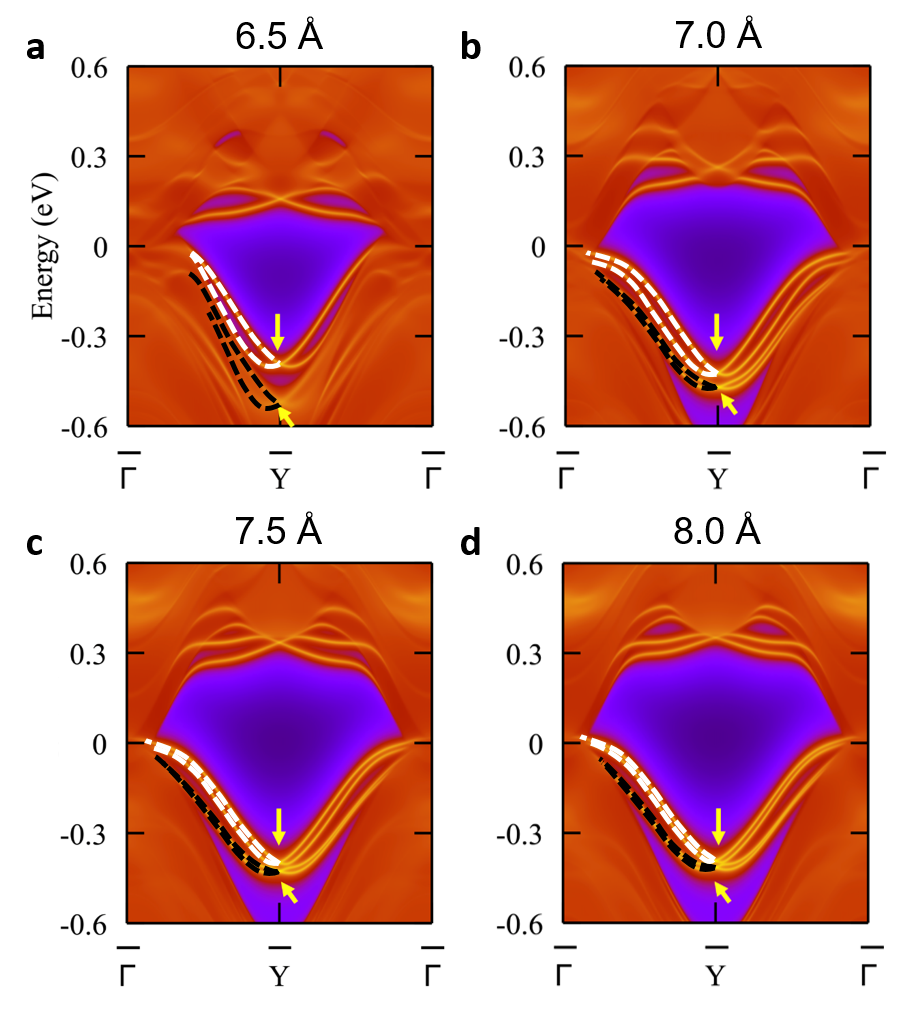}
		\caption{\label{Fig6}\textbf{Dependence of tear-and-stack bilayer edge states on interlayer spacing.} \textbf{(a)-(d)} Band structure of a $\phi=0^{\circ}$ $T'$ bilayer at interlayer distances of $d = 6.5\,$\AA\ to $d = 8.0\,$\AA. As the layers are moved further apart from each other, their QSH edge states become increasingly closer in energy due to a reducing charge transfer between the layers. When the layers are completely decoupled, the edge states are identical to that of \WTe ML. Black and white dashed lines in the left half of each panel indicate the edge states in each of the two layers, respectively. Arrows indicate the respective Dirac points.
		}
	\end{figure}
	
	To understand the influence of interlayer distance and stacking on the electronic and topological properties of the \WTe bilayers, we performed density functional theory (DFT) calculations of a \WTe monolayer, trivial bilayer ($\phi=180^{\circ}$), as well as a bilayer with $\phi=0^{\circ}$, most closely resembling our experiments on small twist angle tBLs.
	{In general, we find that weakly coupled \WTe bilayers 
		can be regarded as two decoupled monolayers, each retaining the electronic character of a \WTe monolayer, i.e. the QSH state ($\mathbb{Z}_2=1$).}
	When bringing the layers closer together, they begin to interact and hybridize, leading to a global trivial topology, $\mathbb{Z}_2$ = 0, in agreement with our experiments and earlier studies \cite{Muechler2016}.
	The resulting charge transfer between the layers further leads to an increasing shift of their edge states with respect to each other (Fig. \ref{Fig6}).
	{We further investigate the effect of the subtle structural differences between the 1T' and the 1T$_d$ phase on our calculations (see section S9) and find that the electronic and topological properties are strongly dependent on the stacking geometry owing to their different symmetries and interlayer spacings, which result in different amounts of charge transfer between the layers (Figs. S21 and S22).
		Interestingly, as we bring the bilayers very close together ($d\approx7.25\,$\AA\ in AA stacking), our calculations predict another topological transition back to $\mathbb{Z}_2$  = 1 for the T$_d$ phase (section S9 and Fig. S17).}
	This interlayer distance is however significantly smaller than the $d\approx8\,$\AA\ realized in the AA location of our $\phi=5.5^{\circ}$ sample.
	
	In summary, we conclude that it is possible to tune the topology of \WTe bilayers via their twist angle and interlayer interactions.
	The discovered twist angle-dependent charge-transfer further makes tBL \WTe a suitable system to probe the QSH edge-states of twisted bilayers in transport measurement since bulk conduction can be suppressed to not obscure the edge state transport.

	\section*{Materials and Methods}
	\subsection*{Sample preparation}
	WTe$_2$ and MoS$_2$ were exfoliated onto $285\rm\,nm$ SiO$_2$/Si substrates.
	Cleaving WTe$_2$ crystals produces flakes which have a high probability of producing edges that are aligned with one of the two crystal axes.
	We picked up a $(20\pm1)\rm\,nm$-thick MoS$_2$ flake with a PPC/PDMS stamp and subsequently a WTe$_2$ flake containing mono- and bilayer regions. 
	After the WTe$_2$ pickup, the heterostructure was flipped upside down and placed onto a gold lead on an SiO$_2$ substrate which was pre-mounted and contacted to an STM sample plate. 
	The tear-and-stack sample was fabricated in the same way, except that two pick-ups of the \WTe took place with a slight rotation of the substrate after the first pickup. 
	The exfoliation and stacking took place in a nitrogen-filled glovebox, from which the samples were transferred into a high vacuum tube furnace to remove the PPC by annealing at $250^{\circ}\rm C$ for $8\rm\,h$. Subsequently, the samples are transferred into the STM ultra-high vacuum chamber where a final anneal at $250^{\circ}\rm C$ for $\SI{\sim10}{min}$ ($p\leq\SI{1e-8}{mbar}$) was performed, all sample transfers took place in nitrogen atmosphere. 
	
	\subsection*{STM measurements}
	{STM measurements are performed using a commercial CreaTec setup}  with a sample stage temperature of $\SI{4.7}{K}$.
	Electrochemically-etched tungsten tips were indented and checked for a clean spectrum on the gold leads prior and in between measurements. 
	The measurements throughout the manuscript have been performed with multiple iterations of freshly prepared tip apices to exclude tip artifacts. 
	$dI/dV$ measurements were performed using a lock-in amplifier set to a frequency of $f=869\rm\,Hz$, at stabilization setpoint $V_{\rm sample} = 500\rm\,mV$, $I_{\rm t} = 50\rm\,pA$ and with a modulation amplitude $V_{\rm mod} = 10\rm\,mV$, except for Fig. 1d and 4c, for which we used $V_{\rm sample} = 1\rm\,V$, $I_{\rm t} = 100\rm\,pA$, with a modulation amplitude $V_{\rm mod}=20\rm\,mV$
	
	\section*{Acknowledgments}
	\textbf{Funding:} We acknowledge helpful discussions with Justin Song. P. G. would like to acknowledge fruitful discussions with Jaron T. Krogel and Yubo `Paul' Yang.  B.M.H. was supported by the Department of Energy under the Early Career award program (\#DE-SC0018115).
	F.L. and D.W. were supported by the NSF DMR-1809145 for the STM measurements. The authors gratefully acknowledge NSF DMR-1626099 for acquisition of the STM instrument.
	F.L. was supported by the Center for Nanophase Materials Sciences, Oak Ridge National Laboratory, which is a DOE Office of Science User Facility.
	F.L. acknowledges funding from the Alexander von Humboldt foundation through a Feodor Lynen postdoctoral fellowship.
	F.L. further acknowledges funding by the Deutsche Forschungsgemeinschaft (DFG, German Research Foundation) within the Priority Programme SPP 2244 “2DMP”
	and
	the Bavarian Ministry of Economic Affairs, Regional Development and Energy within Bavaria’s High-Tech Agenda Project ''Bausteine f\"ur das Quantencomputing auf Basis topologischer Materialien mit experimentellen und theoretischen Ans\"atzen''.
	The authors thank the Pennsylvania State University Two-Dimensional Crystal Consortium - Materials Innovation Platform (2DCC-MIP), which is supported by NSF DMR-1539916 for supplying further 2D materials. 
	While support for initial work by A.D.P. and P. G. was by the Oak Ridge National Laboratory’s Laboratory Directed Research and Development project (Project ID 7448, PI: P.G.), work on the key final results presented in the manuscript was supported by the U.S. Department of Energy, Office of Science, Basic Energy Sciences, Materials
	Sciences and Engineering Division, as part of the Computational Materials Sciences Program and Center for Predictive Simulation of Functional Materials. Computations were performed on the Compute and
	Data Environment for Science (CADES) cluster at the Oak Ridge National Laboratory, which is supported by the Office of Science of the U.S. Department of Energy under Contract No. DE-AC05-00OR22725, at the Center for Nanophase Materials Sciences, Oak Ridge National Laboratory, which is a DOE Office of Science User Facility. Crystal growth and characterization at ORNL was supported by the US Department of Energy, Office of Science, Basic Energy Sciences, Division of Materials Sciences and Engineering.
	\textbf{Author contributions:} F.L. and D.W. fabricated the samples, measured and analyzed the experimental data and wrote the manuscript. P.G. and A. D. P. performed the DFT calculations, interpreted the results and contributed to writing the manuscript. J.Y. and D.G.M. grew the \WTe crystals. B.M.H. supervised the project. All authors commented on the manuscript.
	\textbf{Competing interests:} The authors declare that they have no competing interests. 
	\textbf{Data and materials availability:} All data needed to evaluate the conclusions in the paper are present in the paper and/or the Supplementary Materials. Additional data related to this paper may be requested from the authors. 
	
	\bibliographystyle{naturemag}

	\clearpage

	
	

	\clearpage
	
	\onecolumngrid 
	\cleardoublepage
	\setcounter{figure}{0}
	\setcounter{equation}{0}
	\setcounter{equation}{0}
	\setcounter{mybibstartvalue}{27}
	
	\renewcommand\thesection{S\arabic{section}}
	\renewcommand\thesubsection{S\thesection.\arabic{subsection}}
	\renewcommand{\thefigure}{S\arabic{figure}}
	
	\renewcommand{\theequation}{S\arabic{equation}}
	\renewcommand{\thetable}{S\arabic{table}}
	
	\makeatletter
	\renewcommand\@biblabel[1]{[S#1]} 
	\makeatother
	\renewcommand{\citenumfont}{S}
	
	\begin{center}
		{\textbf{\large Supplementary Information: Quantum spin Hall edge states and interlayer coupling in twisted-bilayer 
				\texorpdfstring{\WTe}{WTe2}\\}}
		\vspace{0.5cm}
		{Felix L\"upke$^{1, 2, 3, 4}$, Dacen Waters$^{1, 5}$, Anh D. Pham$^{2}$, Jiaqiang Yan$^{6}$, \\ David G. Mandrus,$^{3,6,7}$, Panchapakesan Ganesh$^{2}$,
			Benjamin M. Hunt$^{1, \dag}$}
		\vspace{0.3cm}
		
		\normalsize{$^{1}$Department of Physics, Carnegie Mellon University, Pittsburgh, PA 15213, USA}\\
		\normalsize{$^{2}$Center for Nanophase Materials Sciences, Oak Ridge National Laboratory, \\Oak Ridge, TN 37831, USA}\\
		\normalsize{$^{3}$Department of Materials Science and Engineering, University of Tennessee, Knoxville, TN 37996, USA}\\
		\normalsize{$^{4}$Peter Gr\"unberg Institut (PGI-3), Forschungszentrum J\"ulich, 52425 J\"ulich, Germany}\\
		\normalsize{$^{5}$Department of Physics, University of Washington, Seattle, WA 98195, USA}\\
		\normalsize{$^{6}$Materials Science and Technology Division, Oak Ridge National Laboratory, \\Oak Ridge, TN 37831, USA}\\
		\normalsize{$^{7}${Department of Physics and Astronomy, University of Tennessee, Knoxville, TN 37996, USA}}\\
		\normalsize{$^{\dag}$ E-mail: bmhunt@andrew.cmu.edu}
		\vspace{0.5cm}
	\end{center}

	\section{Folded bilayer sample}
\begin{figure*}[ht]
		\centering
		\includegraphics[width=12cm]{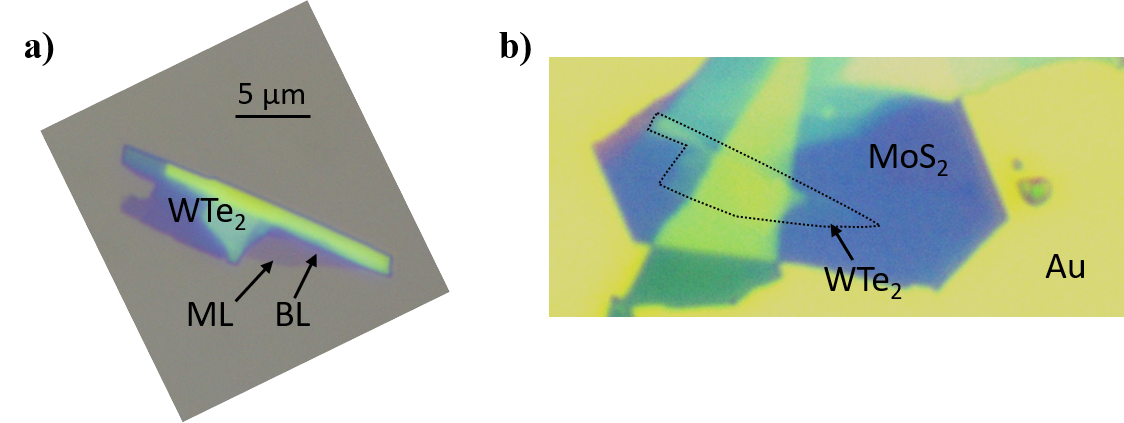}
		\caption{\label{Fig_optical} {Optical microscope images of sample.} {{(a)} Exfoliated WTe$_2$ flake. {(b)} Assembled WTe$_2$/MoS$_2$ heterostructure on Au electrode after overnight annealing.} 
		}
	\end{figure*}

	Data shown in Figs. 1 and 2 of the main text was taken in different regions of the same \WTe/\MoS heterostructure (sample \#1). This heterostructure was fabricated using the dry-transfer flip technique reported in Ref. \cite{SI-Lupke2019}. Flakes are exfoliated from bulk MoS$_2$ and WTe$_2$, which was characterized to be in the 1T' phase, onto 285 nm SiO$_2$. After exfoliation, the \MoS flake is picked up with a PPC/PDMS stamp, followed by the \WTe flake. Subsequently, the  heterostructure is flipped upside down and then placed on a Cr/Pd/Au lead which was pre-evaporated on SiO$_2$ using optical lithography and is connected to a contact of the STM sample plate.
	The sample is then transferred to a vacuum annealer ($p<1\times10^{-5}\rm\,mbar$) and annealed over night. From there, the sample is transferred back to the glovebox to take optical images and then to the STM chamber. For all transfers, the sample is always in a nitrogen atmosphere and never exposed to air. 
	The sample fabrication took less than $24\rm\,h$ starting from exfoliation in the nitrogen-filled glove box (O$_2$, H$_2$O $< 5\rm\, ppm$).
	In our experience, the \MoS thickness has to be $\lesssim 100\rm\,nm$ in order to be able to tunnel on it at $\sim4\rm\,K$, because otherwise it is too insulating. 
	Optical images of the exfoliated \WTe flake on $285\rm\,nm$ SiO$_2$ and of the assembled heterostructure are shown in Fig. \ref{Fig_optical}.
	
	\begin{figure*}[ht]
		\centering
		\includegraphics[width=14cm]{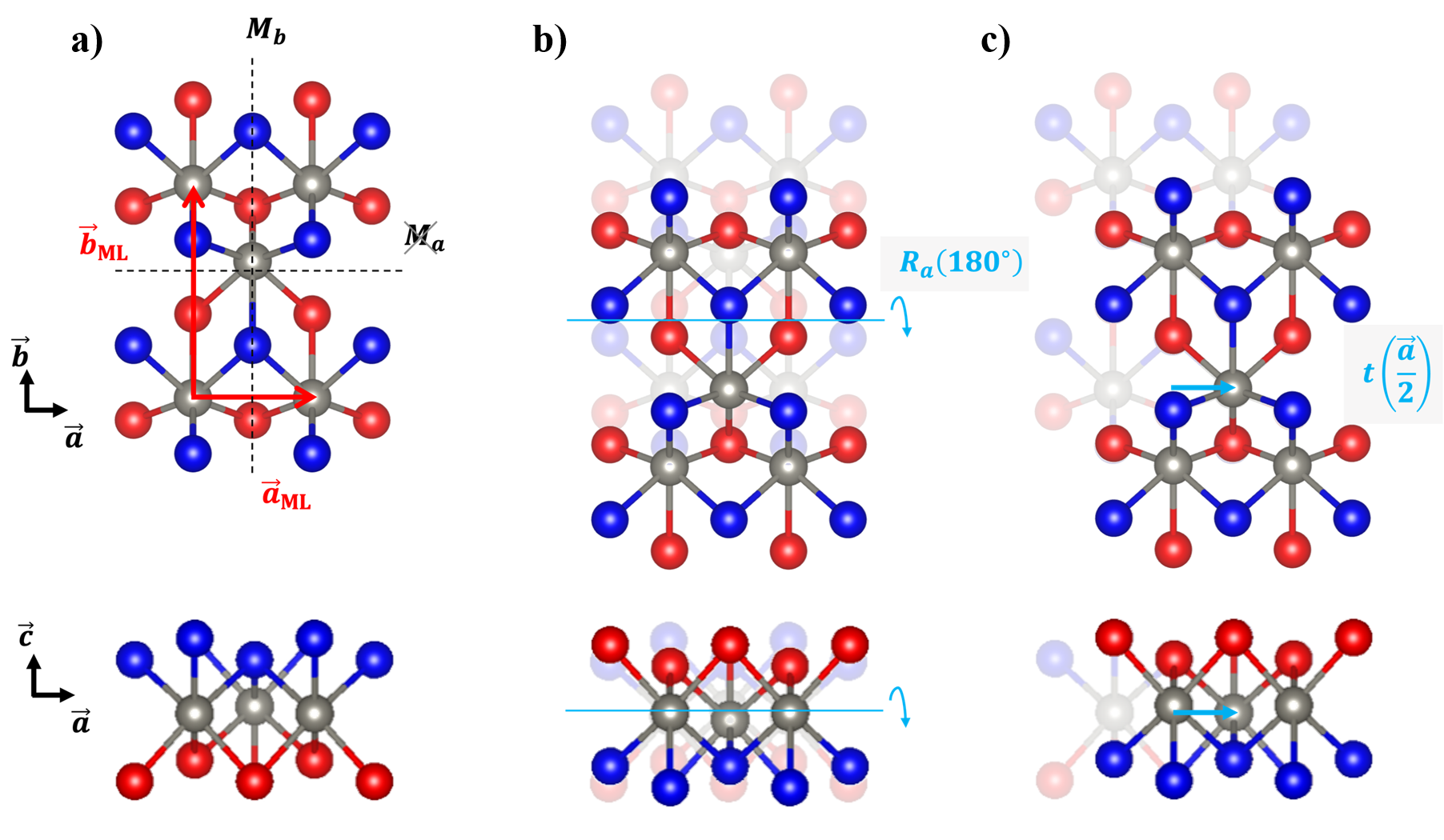}
		\caption{\label{Fig_c2a} {Depiction of the $\bar{C}_{2a}$ symmetry of 1T'-\WTe. Top and side views of the crystal structure, with the top (bottom) layer Te atoms in blue (red). (a) Monolayer unit cell vectors denoted in red, with mirror plane symmetry across the $b$-$c$ plane denoted as $M_b$. There is no mirror symmetry across the $a$-$c$ plane ($M_a$). The original structure is recovered after a rotation of $180^\circ$ about the indicated $a$-axis (panel (b)) and a translation by half a unit cell along the same axis (panel (c)). The original unit cell is shown with reduced opacity in b and c.} 
		}
	\end{figure*} 
	
	\begin{figure*}[h]
		\centering
		\includegraphics[width=14cm]{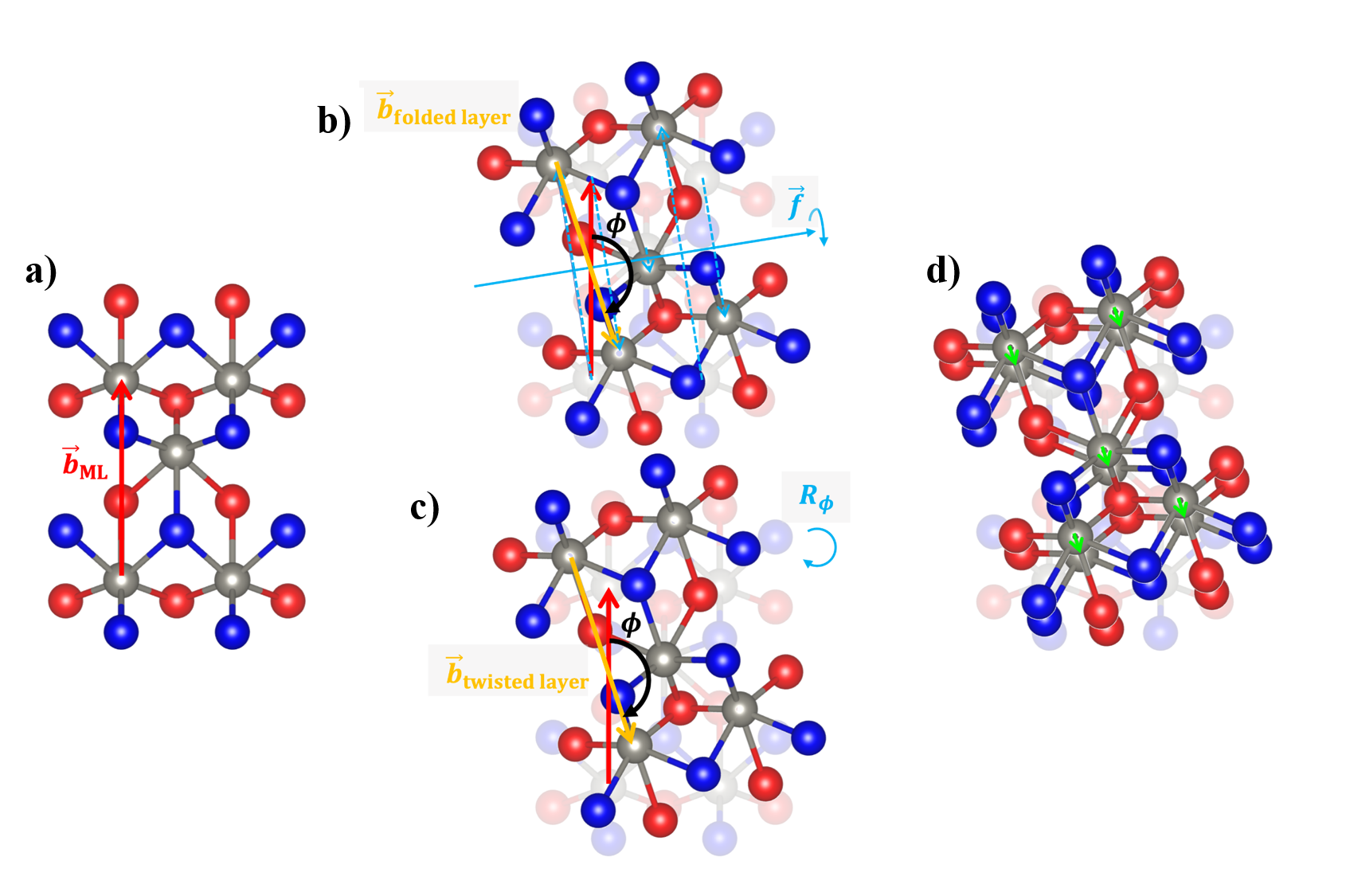}
		\caption{\label{Fig_twist_fold} {(a) Monolayer 1T'-\WTe, with lattice vector $b$ noted. Coloring of the Te atoms are the same as in Fig. \ref{Fig_c2a}. (b) Rotation operation about an arbitrary axis $\vec{f}$, representing a fold like what occurs in our STM samples. The dashed blue arrows track the positions of the W atoms with the fold operation. The orange vector is the $b$-axis of the new folded unit cell. The original unit cell is shown with reduced opacity for comparison, along with the original $b$-axis of the ML unit cell. A twist angle $\phi$ is created. (c) Rotation about the central metal atom by the same angle $\phi$ as in b. (d) The folded (from panel b) and twisted (from panel c) layers, overlaid on the original ML unit cell. The folded and twisted cases are equivalent, modulo a translation depicted by the green arrows.} 
		}
	\end{figure*}

	\section{Crystal symmetry and folded/twisted bilayer comparison}
	
	1T'-\WTe has a screw-axis symmetry along the $a$-axis, i.e. $\bar{C}_{2a}$. The screw transformation is demonstrated stepwise in Fig. \ref{Fig_c2a}. Rotation about the $a$-axis by $180^\circ$ and a translation by half a unit cell along the $a$-axis returns the original unit cell.
	The screw operation resembles a fold, as in the folded bilayers presented in the main text. As a result of the symmetry, a folded layer bilayer can be considered equivalent to a twisted bilayer, as demonstrated in Fig. \ref{Fig_twist_fold}. Folding along an arbitrary axis $\hat{f}$ (i.e. a $180^\circ$ rotation about that axis, as shown in Fig. \ref{Fig_twist_fold}b) results in a rotation $\phi$ between the folded unit cell vectors and the original unit cell vectors. Similarly, by just rotating about the $c$-axis (as in Fig. \ref{Fig_twist_fold}c), the same twist angle $\phi$ can be obtained. It follows that the two scenarios are equivalent, modulo an arbitrary lateral translation (Fig. \ref{Fig_twist_fold}d).
	
	\begin{figure*}[ht]
		\centering
		\includegraphics[width=8cm]{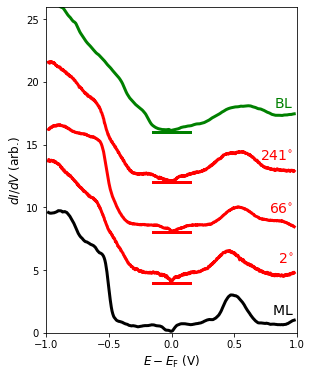}
		\caption{\label{Fig_Spectra} {Summary of tunneling spectra acquired on the different regions of the folded \WTe sample.} 
		}
	\end{figure*} 
	
	\section{Summary of tunneling spectra}
	Tunneling spectra in the range $\pm1\rm\,V$ acquired on the different regions of the \WTe flake of sample \#1 are shown in Fig. \ref{Fig_Spectra}. In as-grown bilayer \WTe, the hybridization of the two layers leads to a semi-metallic state which is evident in the spectra as an increased density of states especially on the valence band side. This increase is mostly absent in the folded bilayers. 
	{
		\section{Additional folded bilayers}
		\begin{figure*}
			\centering
			\includegraphics[width=\textwidth]{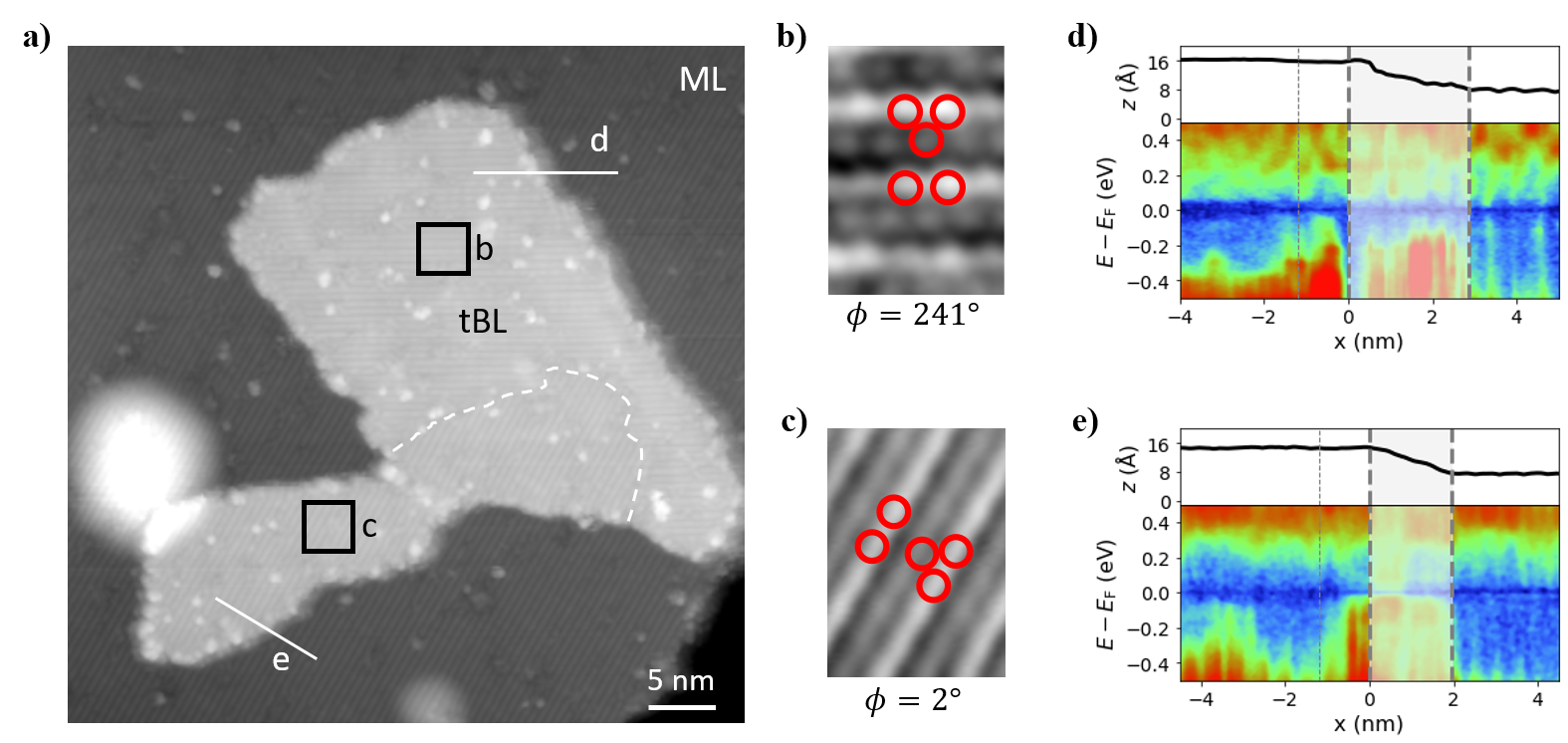}
			\caption{\label{Another_Fold} {Observation of edge states for additional tBL twist angles.} {(a)} Large area scan of an additional folded \WTe monolayer region with a domain boundary (indicated by dashed line, see Supplementary Section 3 for details).
				{(b, c)} Atomically resolved scans on the two tBL \WTe regions indicate a rotational misalignment relative to the monolayer of $241^{\circ}$ and $2^{\circ}$, respectively.
				{(d, e)} Tunneling conductance across the tBL-ML \WTe step edge in the two different regions indicates the presence of an edge state for both tBL regions. Dashed black lines and greyed-out region indicate the same $\sim1.2\rm\,nm$ spatial extent of the QSH edge state and tip/edge artifact region, respectively, as in Fig. 2 of the main text.
			}
		\end{figure*} 
		We have analyzed another folded region  of sample \#1 which has two domains, separated by a lateral domain boundary (Fig. \ref{Another_Fold}a). From atomic resolution scans of the two different domains (Fig. \ref{Another_Fold}b and c), we find that $\phi=241^\circ$ and $\phi=2^\circ$, respectively. Tunneling conductance measurements across the edge of the two additional tBL regions (Fig. \ref{Another_Fold}d and e) show similar results as in Fig. 2 of the main text. 
		While unsurprising for the large twist angle of $\phi=241^{\circ}$, for $\phi=2^{\circ}$ it is somewhat suprising that the electronic structure in this domain still represents that of a decoupled monolayer (for spectra in $\pm1\rm\,V$ range see Fig. \ref{Fig_Spectra}), as in other tBL TMDs or graphene, highly correlated phenomena are reported at similar twist angles \cite{SI-Cao2018_SC, SI-wang2019magic}.
		We explain this observation by the fact that an important aspect of the correlated phenomena in magic-angle bilayers are flat bands which are formed due to the periodicity of the resulting moir\e pattern.
		However, for $\phi=2^\circ$ we estimate the \WTe moir\e pattern periodicity to be
		$L\approx b/\sqrt{\varepsilon^2 + \phi^2}=18\,\rm nm$, where $b=0.628\,\rm nm$ and the lattice mismatch $\varepsilon=0$, i.e. comparable to the size of the entire domain. 
		We therefore conclude that moir\e effects do not play a role in this domain despite the small twist angle. 
		We have further performed tunneling conductance measurements across the lateral domain boundary which agree well with domain boundaries in epitaxially grown 1T'-WSe$_2$ monolayers \cite{SI-Pedramrazi19} as discussed below.
	}
	
	\section{Topologically trivial domain boundary state}
	\begin{figure*}[ht]
		\centering
		\includegraphics[width=\textwidth]{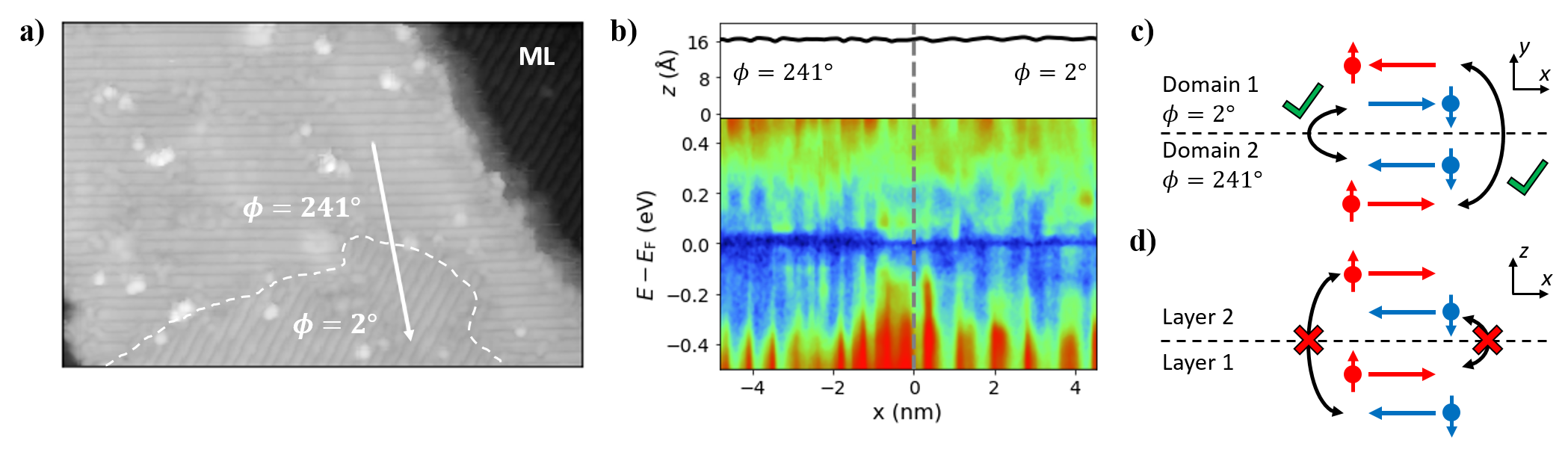}
		\caption{\label{Fig_Domain} {Topologically trivial domain boundary states.} {(a)} Zoomed in topography of the domain boundary in Fig. 3 of the main text. 
			{(b)} Tunneling conductance across the domain boundary, along the arrow indicated in (a) showing the presence of a boundary state.
			{(c)} Schematic of the counter-propagating QSH edge states at the domain boundary. Due to the relative orientation of the QSH edge states along the domain boundary, backscattering is possible.
			{(d)} Schematic of the two sets of QSH edge states sitting on top of each other at the tBL-\MoS edge shown in the main text, i.e. along the edge where the fold occured. 
			Spin-momentum locking of the edge states in combination with the \WTe monolayer symmetry and weak interlayer coupling suppress intra- and inter-layer backscattering. 
		}
	\end{figure*}  
	Figure \ref{Fig_Domain}a shows a zoom into the domain boundary on the folded \WTe layer shown in Fig. \ref{Fig_Domain}. 
	The relative crystallographic orientation between the two domains corresponds to $120^\circ(\approx360^\circ - (241^\circ - 2^\circ))$ withing measurement error (Fig. \ref{Fig_Domain}a).
	While the observation of domain boundaries in exfoliated flakes is rare in our experience, the $120^\circ$ domain boundary was reported in epitaxially grown 1T'-WSe$_2$ as the most common configuration \cite{SI-Pedramrazi19}.
	While detailed measurements at the domain boundary are difficult because of its irregular shape and many defects along the boundary, Fig. \ref{Fig_Domain}b shows a tunneling conductance measurement across the domain boundary in which we observe an increased density of states at the position of the domain boundary, similar to that of the QSH edge state feature. This observation is in agreement with the presence of two instances of counter-propagating QSH edge states, one on each side of the domain boundary.
	However, in this configuration backscattering from one QSH channel into the other is possible (Fig. \ref{Fig_Domain}c) and the channels hybridize. This is in contrast to the configuration of two QSH edge states sitting on top of each other in the folded bilayers as shown Fig. 2 of the main text.
	The observation of counter-propagating QSH states at the domain boundary supports our conclusion that both, the $\phi=241^{\circ}$ and $\phi=2^{\circ}$ domain, are topologically non-trivial. These observations are consistent with previous work on domain boundary states in epitaxially grown 1T'-WSe$_2$ \cite{SI-Pedramrazi19}.

	\section{Edge detection scheme}
\begin{figure*}
		\centering
		\includegraphics[width=16cm]{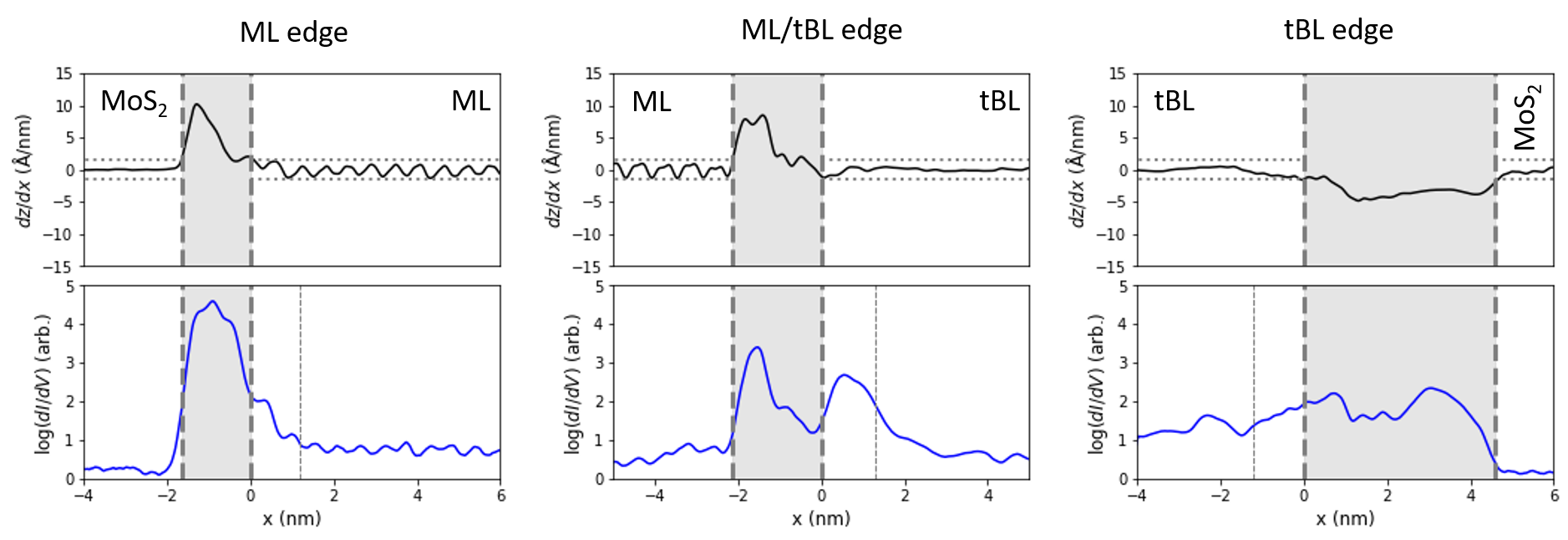}
		\caption{\label{Fig_Edge_detection_1} Edge detection for Fig. 2 of the main text.
			(Top row) Topography gradient with the threshold for edge detection $dz/dx=1.5\,\rm\AA/nm$ indicated as horizontal lines indicate. Dashed vertical lines indicated the resulting detected edge positions.
			(Bottom row) Corresponding $dI/dV$ signal across the step edge, i.e. cross sections of the $dI/dV$ maps at $E-E_{\rm F}=-300\rm\,meV$ in the corresponding figures.
			Thin dotted vertical lines indicate the extension of the edge state from the edge according to the literature value of 1.2 nm.
		}
	\end{figure*}

\begin{figure*}
		\centering
		\includegraphics[width=16cm]{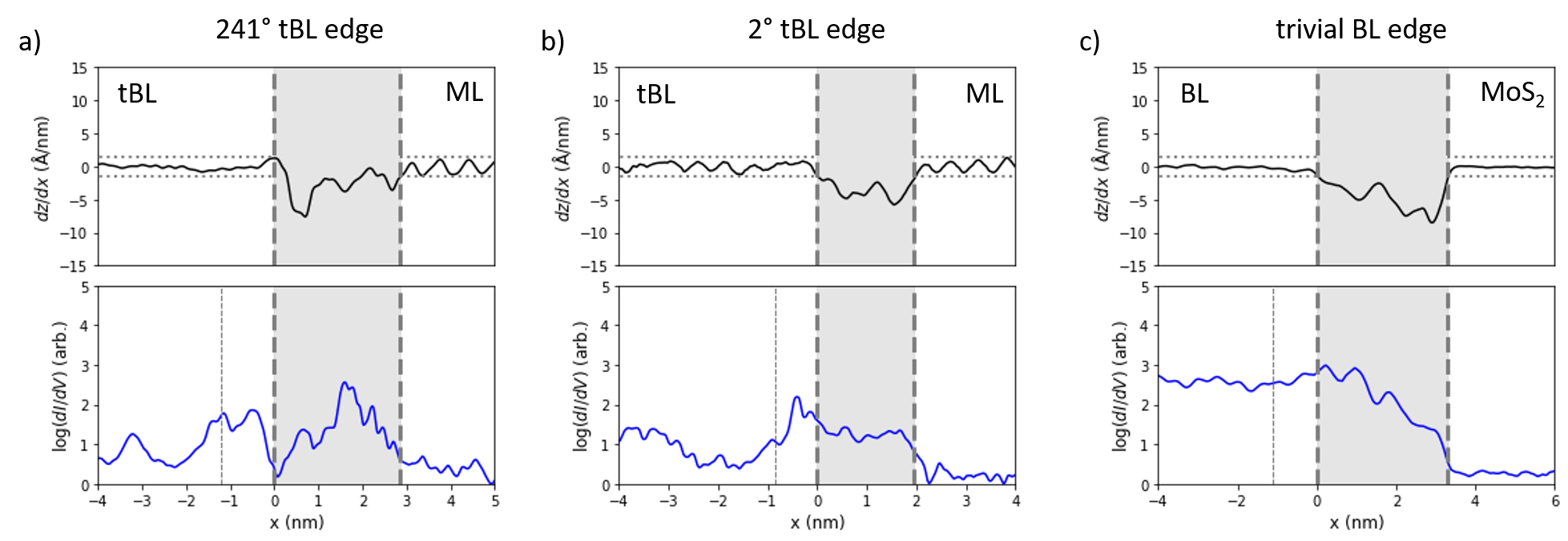}
		\caption{\label{Fig_Edge_detection_2} Edge detection for tBL edges for (a) Fig. S3d, (b) Fig. S3e, (c) Fig. 3d of the main text. 
			(Top row) Topography gradient with the threshold for edge detection $dz/dx=1.5\,\rm\AA/nm$ indicated as horizontal lines indicate. Dashed vertical lines indicated the resulting detected edge positions.
			(Bottom row) Corresponding $dI/dV$ signal across the step edge, i.e. cross sections of the $dI/dV$ maps at $E-E_{\rm F}=-300\rm\,meV$ in the corresponding figures.
			Thin dotted vertical lines indicate the extension of the edge state from the edge according to the literature value of 1.2 nm.}
	\end{figure*}

	To identify the positions of the topographic edges, we use an edge detection scheme based on the topography gradient. We apply this scheme to each edge separately, but using the same criteria to achieve an edge detection that does {\it not} require subjective inputs for each data set. To detect the edge positions, we determine a threshold for the topography gradient of $dz/dx=1.5\rm\,\AA/nm$ after applying a moving average filter to achieve a robust detection.
	Approaching an edge, the position of the edge is determined as the position where $dz/dx$ reaches this threshold. Figs. \ref{Fig_Edge_detection_1} and \ref{Fig_Edge_detection_2} show the edge detection for the data sets analyzed in the main paper as well as Fig. \ref{Another_Fold}, with the respective threshold values indicated as horizontal dotted lines, as well as the corresponding $dI/dV$ signal at $E-E_{\rm F}=-300\rm\,meV$.
	

	\section{Exclusion of tip artifacts at ML-BL junction}
\begin{figure*}[!ht]
		\centering
		\includegraphics[width=8cm]{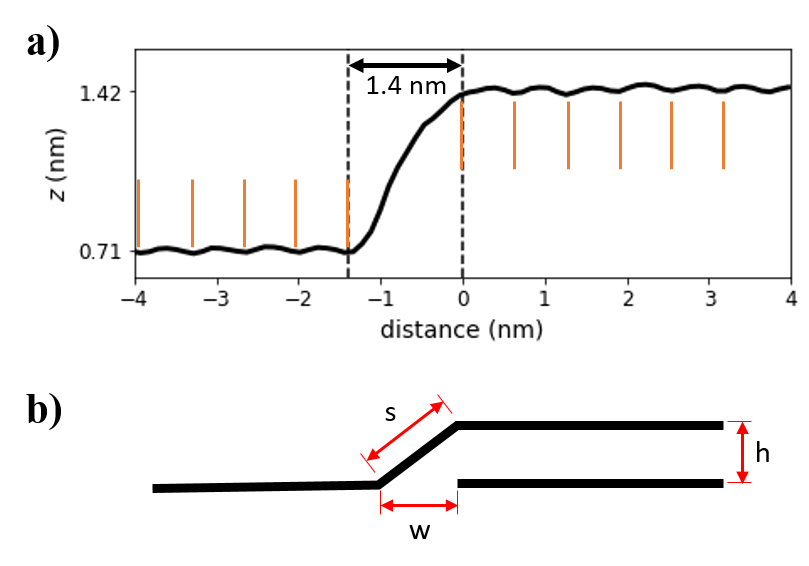}
		\caption{\label{Fig_ML-BL} {Exclusion of tip artifacts at ML-BL junction.} 
			{(a)} Profile across the ML-BL junction. The dashed lines indicate the transition region corresponding to Fig. 4 in the main text. Orange lines indicate the spacing of atomic rows on the each terrace. Dashed lines indicate the transition region which consists of three \WTe unit cells. 
			{(b)} Geometric model of the monolayer draping over the layer below. For $s=3b=1.88\rm\,nm$ and $h=0.71\,\rm nm$, it results $w=1.74\rm\,nm$.
		}
	\end{figure*}

	To confirm the measurement of the continuous \WTe layer draped over a second layer below without artifacts, we carefully analyze the lateral position of the atomic rows across the step edge (Fig. \ref{Fig_ML-BL}a). We find that the transition region corresponds to three \WTe unit cells which corresponds to a length of $s=3b=\SI{1.88}{nm}$.
	In combination with the observed step height of $h=7.1\,$\AA, this allows us to estimate the width of the transition region to be $\sqrt{s^2-h^2}=1.74\rm\,nm$ (Fig. \ref{Fig_ML-BL}b). While this value is slightly greater than the experimentally observed transition region ($w=1.4\rm\,nm$) we attribute this difference to the 's' shape of the transition region which results in a smaller lateral extension compared to our simple model. 
	However, more importantly, we take the fact that the observed transition region is less extended than the estimated size as evidence that there are no significant tip artifacts due to a blunt tip, as this would extend the transition region.
	Furthermore, we observe a small modulation of the profile throughout the transition region which corresponds to the expected atomic rows, also supporting the absence of any significant tip artifacts in this measurement.
	We note that while there are dangling bonds at the ML-BL junction, they are located \textit{underneath} the ML and therefore not expected to significantly contribute to the measured tunneling conductance in the top layer of the the ML-BL transition region.\\

	\section{Tear-and-stack bilayer sample}
	\begin{figure*}
		\centering
		\includegraphics[width=0.7\textwidth]{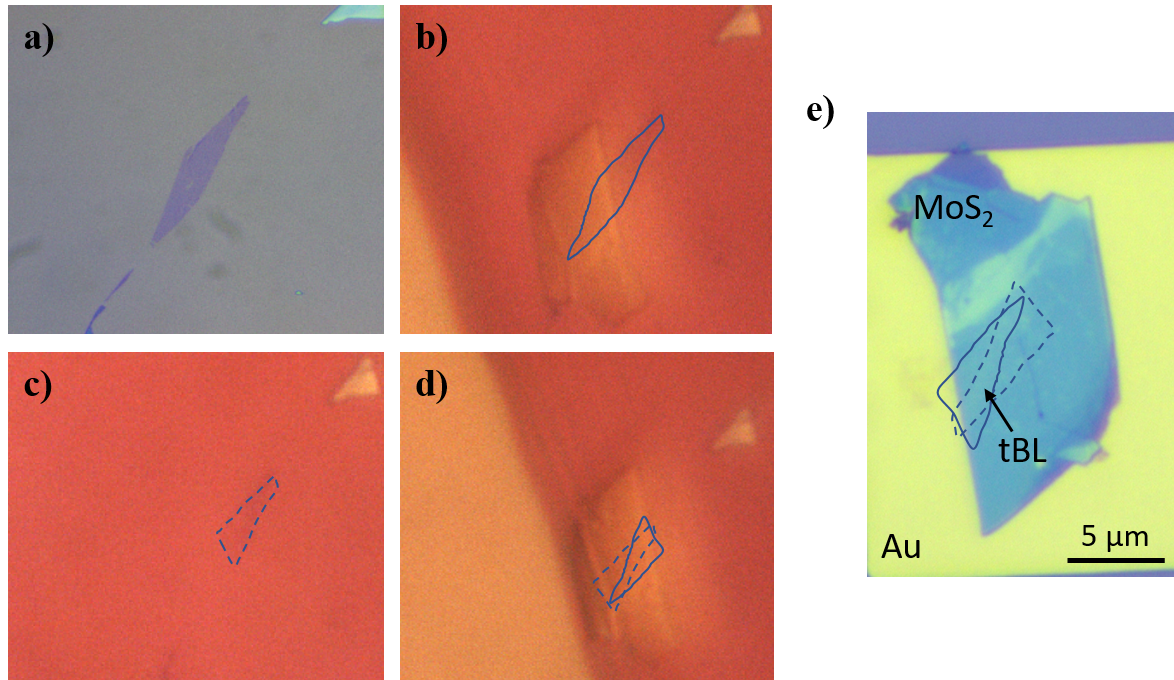}
		\caption{\label{Fig_tBL} {Optical images of tBL sample fabrication} 
			{(a)} Optical image of ML \WTe flake.
			{(b)} Picking up half of the \WTe flake with a \MoS flake (seen through PPC/PDMS stamp). 
			{(c)} After the first pickup, half of the \WTe flake is left on the substrate.
			{(d)} After slight rotation of the transfer stamp, the other half of the \WTe flake is picked up.
			{(e)} Optical image of the heterostructure after flipping, put-down onto a pre-evaporated gold lead and vacuum annealing.
		}
	\end{figure*}   
	\begin{figure*}
		\centering
		\includegraphics[width=\textwidth]{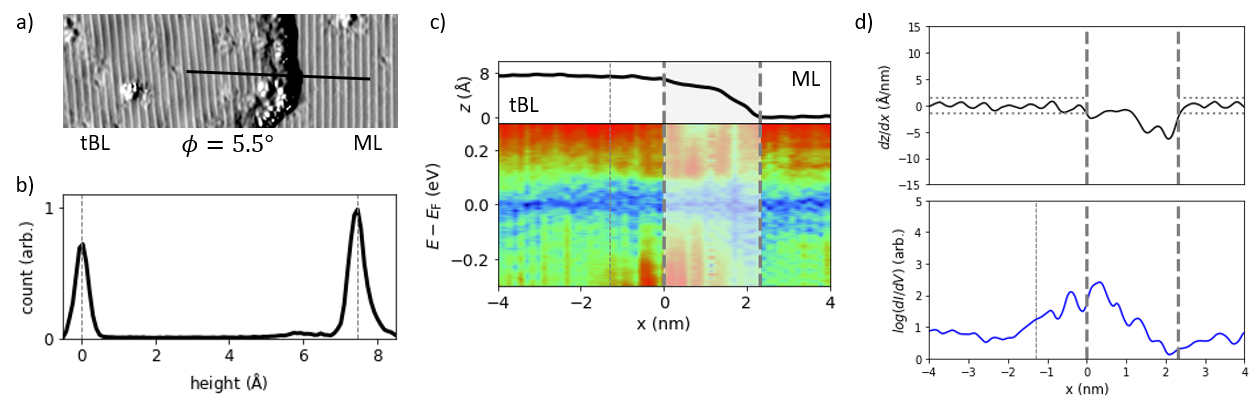}
		\caption{\label{Fig_TAS-tBL} {Edge state in tear-and-stack twisted bilayer \WTe.} 
			{(a)} Gradient image of the step edge from the tBL to the underlying \WTe ML showing the twist angle to be $\phi=5.5^{\circ}$. The corresponding moir\e unit cell length is $L=b/\phi=7.3\rm\,nm$, in agreement with the moir\e period observed in Fig. 4 of the main text.
			{(b)} Tunneling conductance across the ML-tBL junction again showing the presence of an edge state.
			{(c)} Histogram of the tBL step topography (corresponding to (a)) showing an average interlayer distance of 7.45 \AA.
		}
	\end{figure*} 
	\begin{figure*}
		\centering
		\includegraphics[width=16cm]{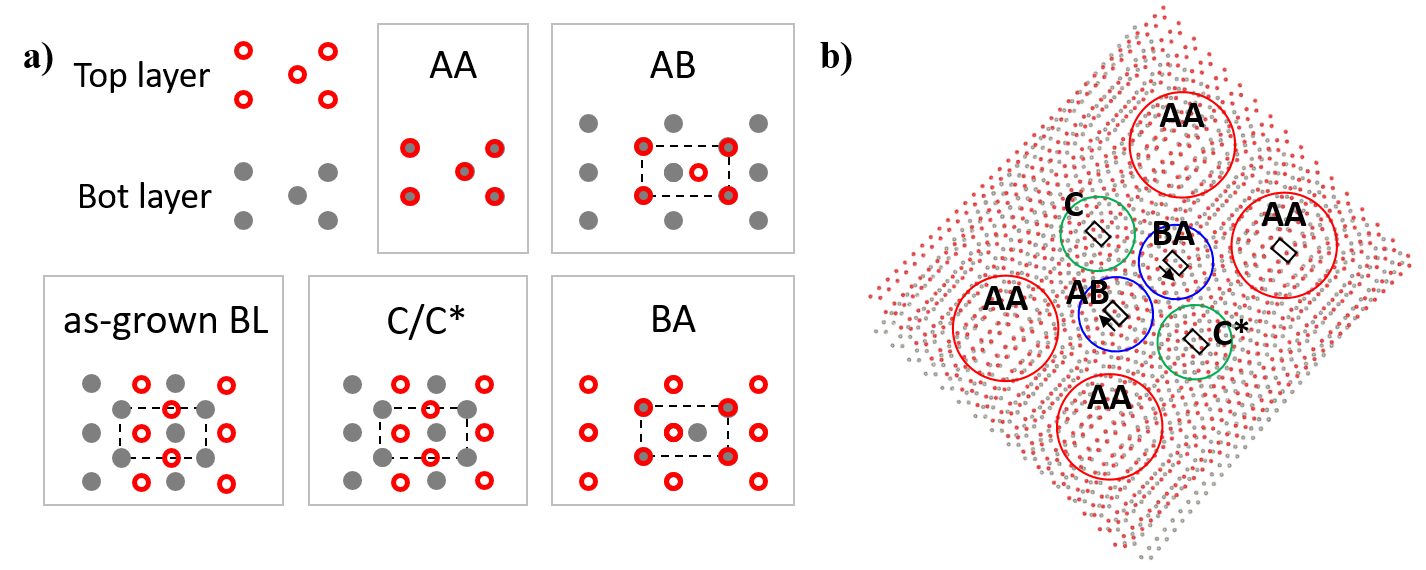}
		\caption{\label{/Fig_TAS_moire} {Atomic model of tear-and-stack twisted bilayer \WTe.}
			(a) Schematic representation of special stacking arrangements in tBL \WTe (only considering the W atoms), as well as as-grown bilayer \WTe as reference. 
			(b) Moir\e pattern formed ($\phi=5.5^\circ$) with the high symmetry points and local stacking arrangements designated. 
		}
	\end{figure*}
	\begin{figure*}[ht]
		\centering
		\includegraphics[width=8cm]{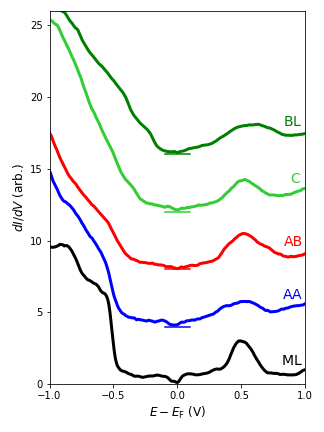}
		\caption{\label{Fig_Spectra2} {Comparison of tunneling spectra acquired in the different regions of the moir\e unit cell of the tear-and-stack sample with monolayer and trivial bilayer spectrum.} 
		}
	\end{figure*} 
	The tear-and-stack tBL \WTe heterostructure (sample \#2) was fabricated utilizing an \MoS substrate flake by first picking up only a portion of the \WTe flake. 
	The substrate + ML \WTe portion was then translated laterally, rotated by a small angle, and aligned to the remaining ML \WTe resulting in a overlap region of about 500 nm width.
	The height difference between the two layers is estimated at the edge of the tBL region to be $7.45\,$\AA\ on average (Fig. \ref{Fig_TAS-tBL}c). At the same edge, we observe an edge state similar to that of a \WTe ML and the folded bilayer sample.
	Atomic resolution images from the ML \WTe region indicate that the twist angle is about $\phi\approx5.5^\circ$. Since the moir\e wavelength is not isotropic, we must consider what interference gives rise to the moir\e pattern, and keep in mind directionality. The direction of the moir\e wavelength $\vec{L}$ can be understood in the context of a beat frequency in real space, formed due to the difference of wavevectors $\vec{k}=\vec{k}_1-\vec{k}_2$. Considering the long-axis of the ML \WTe unit cell as an example (i.e. the real space lattice vector $\vec{b}=b\hat{b}$), we find that for a small twist angle $\phi$, the difference in wave vectors gives $\vec{k}=2\pi b[\hat{a} - (\cos(\phi)\hat{a}+\sin(\phi)\hat{b})]$. For small angles, this is $\vec{k}\approx -2\pi b\hat{b}$ such that the direction of $\vec{L}$ associated with the long-axis moir\e pattern is along the $\hat{a}$ direction. The same logic can be applied to the smaller axis ($a$-axis) of the rectangular ML \WTe unit cell. Therefore, one would expect that there should be a moir\e pattern perpendicular to the long axis of the ML \WTe unit cell of wavelength $\mod{\vec{L}_b}=b/\phi$ and a shorter moir\e pattern perpendicular to the short of axis of the \WTe unit cell of wavelength $\mod{\vec{L}_a}=a/\phi$. The moir\e wavelengths are found to be $\mod{\vec{L}_b}=6.8\pm0.5\ \mathrm{nm}$ and $\mod{\vec{L}_a}=3.5\pm0.3\ \mathrm{nm}$, and the directionality can be seen to align well with our prediction that the long (short) moir\e wavelength is perpendicular to the long (short) axis of the \WTe unit cell (main text, Fig. 5). The angle determined from the moir\e wavelengths are then $\phi=5.3^\circ$ and $\phi=5.5^\circ$, respectively, in excellent agreement with the atomic resolution images. 
	
	To understand the moir\e domains, it is helpful to consider only the metal atoms of the ML \WTe unit cell and that the two layers are rigid lattices (i.e. no relaxation effects \cite{SI-waters2020flatbands}). This is likely a good approximation since the twist angle is relatively large and relaxation effects are not expected to play a significant role. Schematics of a few high-symmetry points in the moir\e pattern between the rotated rigid lattices are shown in Fig. \ref{/Fig_TAS_moire}a. When AA-stacked, all of the W atoms line up, while AB stacking results in one of the W atoms from the top layers covering the other W atom from the bottom layer (and vice versa for BA stacking). 
	A diagram of the moir\e pattern formed from the two layers is shown in Fig. \ref{/Fig_TAS_moire}b. 
	In this picture, AB and BA stacked regions are very close together in the center of the rectangular moir\e unit cell, which might explain why there are not separate domains in STM topography, as evident in Fig. 4 of the main text. The absolute corrugation maxima are likely associated with AA stacking and the second corrugation maxima are associated with the AB and BA stackings. Without relaxation, any topological or spectral differences must be associated with local stacking registries and some small height difference effects due to energetically preferred interlayer distances.
	A comparison of the tunneling spectra acquired in the different moir\e registries is shown in Fig. \ref{Fig_Spectra2}.
	The tear-and-stack fabrication method of this sample caused a relatively discorded edge, limiting the high-quality spectral analysis across the edge state. We include an example line spectra across the edge, indicating a spectral feature similar to the QSH edge state reported in the main text. While overall consistent with our interpretations in the main text, it is difficult to conclude the influence of moir\e pattern on the QSH state due to lack of theoretical predictions as function of twist angle and requirement of higher-quality and smaller-angle devices.\\

	\section{DFT calculations}
	\begin{figure*}
		\centering
		\includegraphics[width=16cm]{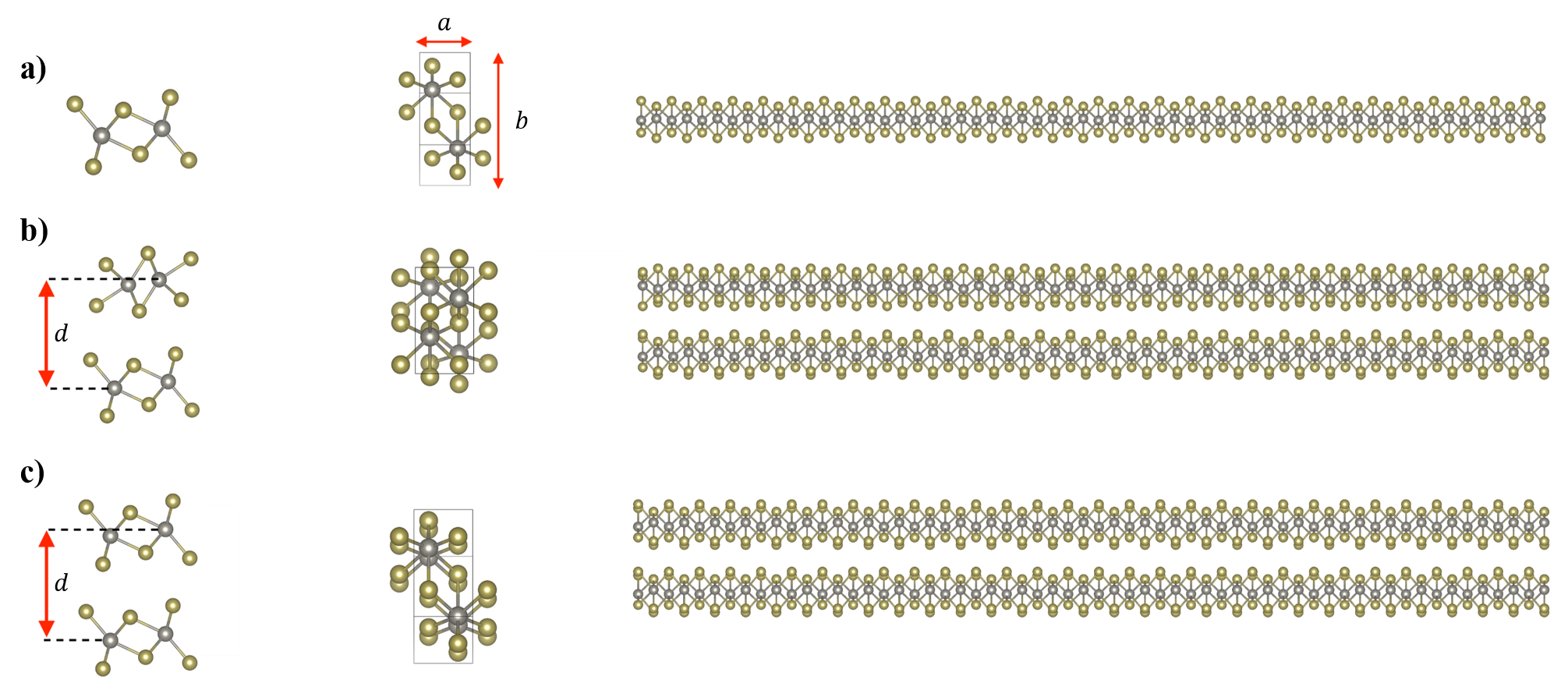}
		\caption{\label{Figs_DFT_2} {Atomic models of WTe$_2$ layers in T' phase.} Bulk and edge atomic structure for {(a)} \WTe monolayer,  
			{(b)} as-grown bilayer ($\phi=180^{\circ}$), and 
			{(c)} $\phi=0^{\circ}$. Here the interlayer distance {\sl d} is defined as the W-W interlayer distance.}
	\end{figure*} 
	\begin{figure*}
		\centering
		\includegraphics[width=16cm]{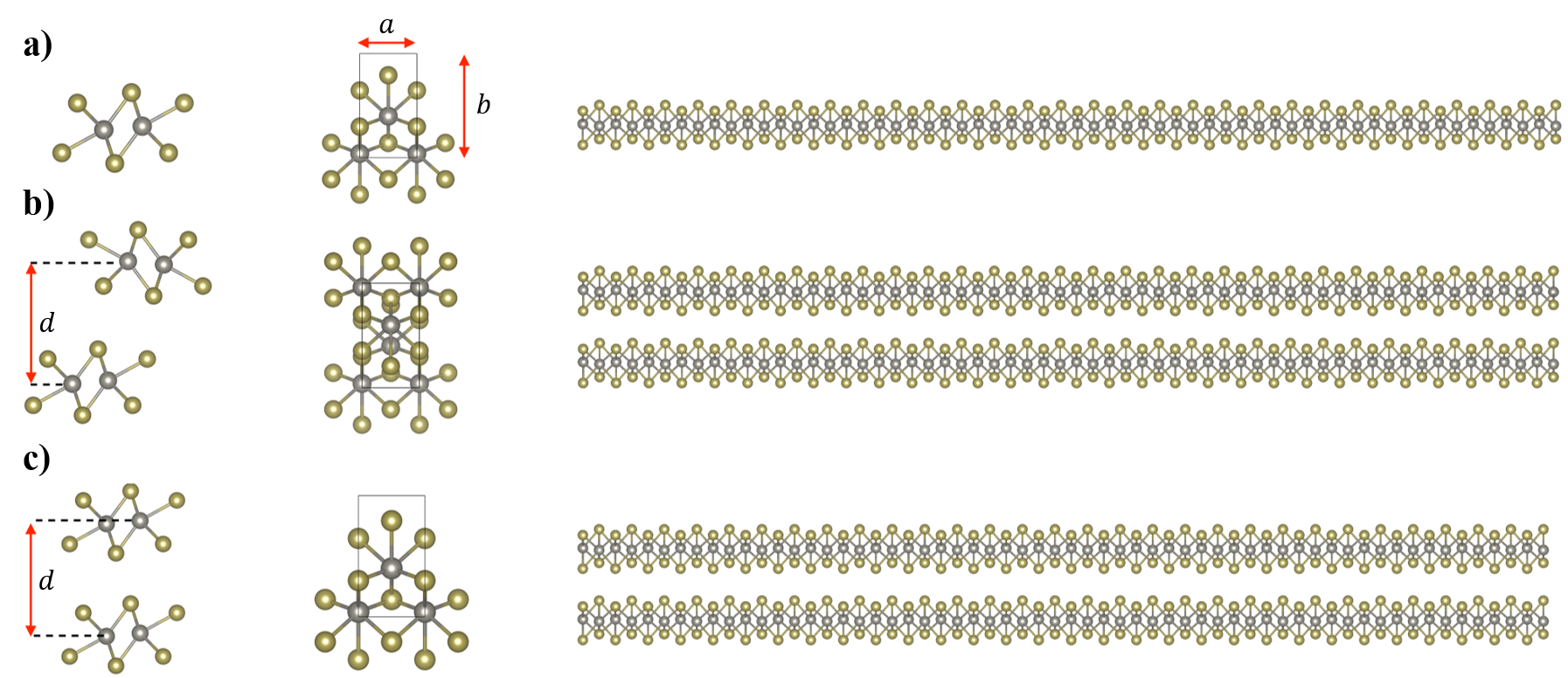}
		\caption{\label{Figs_DFT_1} {Atomic models of WTe$_2$ layers in T$_d$ phase.} Bulk and edge atomic structure for 
			{(a)} \WTe monolayer,  
			{(b)} as-grown bilayer ($\phi=180^{\circ}$), and 
			{(c)} $\phi=0^{\circ}$ (AA-stacking). Here the interlayer distance {\sl d} is defined as the W-W interlayer distance.}
	\end{figure*} 
	\begin{figure*}
		\centering
		\includegraphics[width=\textwidth]{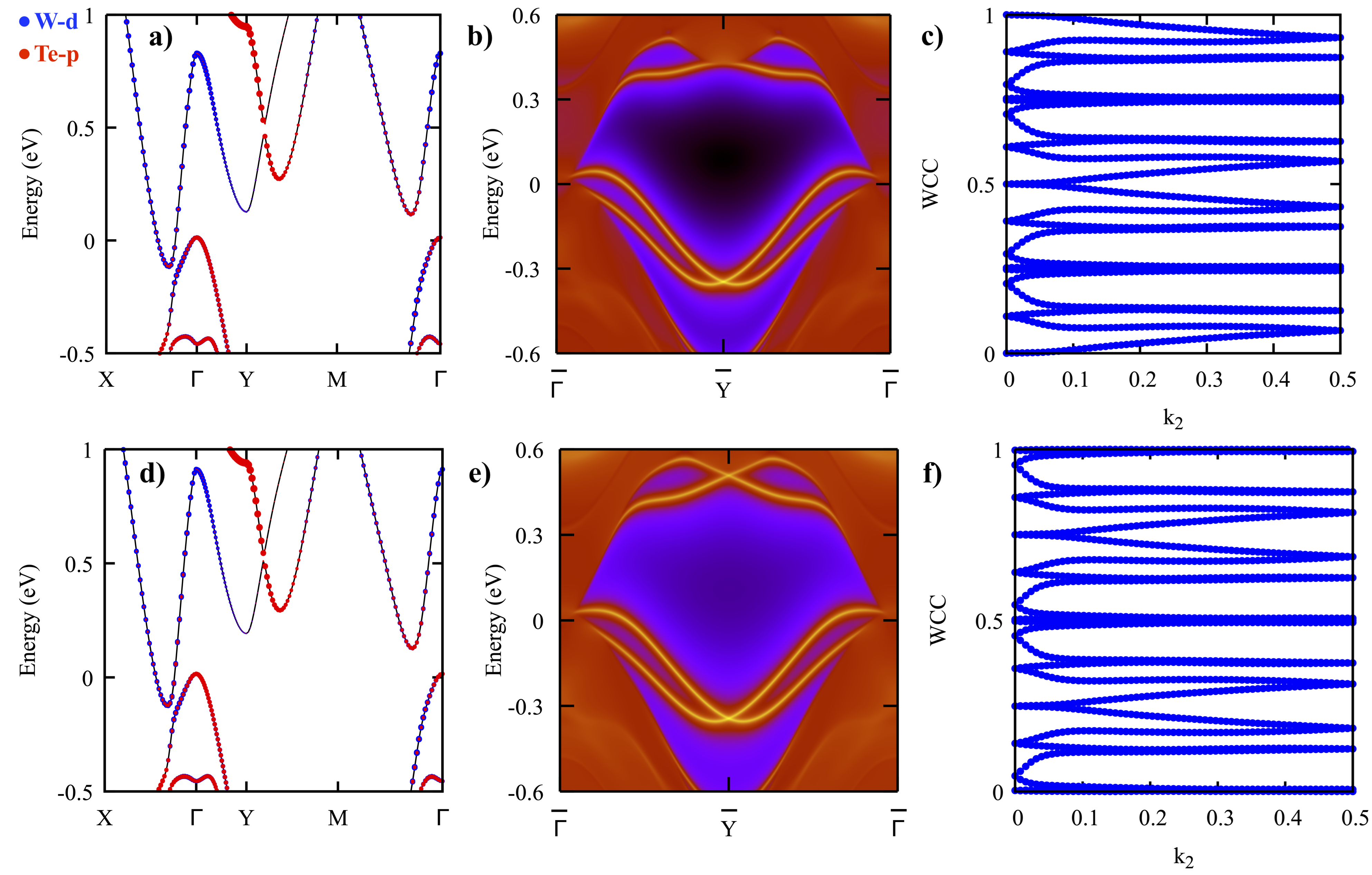}
		\caption{\label{Figs_DFT_3} {Bulk band-structure, edge spectral-functions and Wannier charge center of monolayer WTe$_2$ in T' and T$_d$ phases.}  
			{(a)} Bulk band structure of T' with projected atomic weights of W's {\sl d} and Te's {\sl p} , 
			{(b)} show the spectral function of (010) edges, 
			{(c)} Wannier charge center of T' phase indicating a $\mathbb{Z}_2$ = 1,  
			{(d)} Bulk band structure of T$_d$ with projected atomic weights of W's {\sl d} and Te's {\sl p}, 
			{(e)} show the spectral function of (010) edges, 
			{(f)} Wannier charge center of T$_d$ phase indicating a $\mathbb{Z}_2$ = 1.} 
	\end{figure*} 
	\begin{figure*}
		\centering
		\includegraphics[width=16cm]{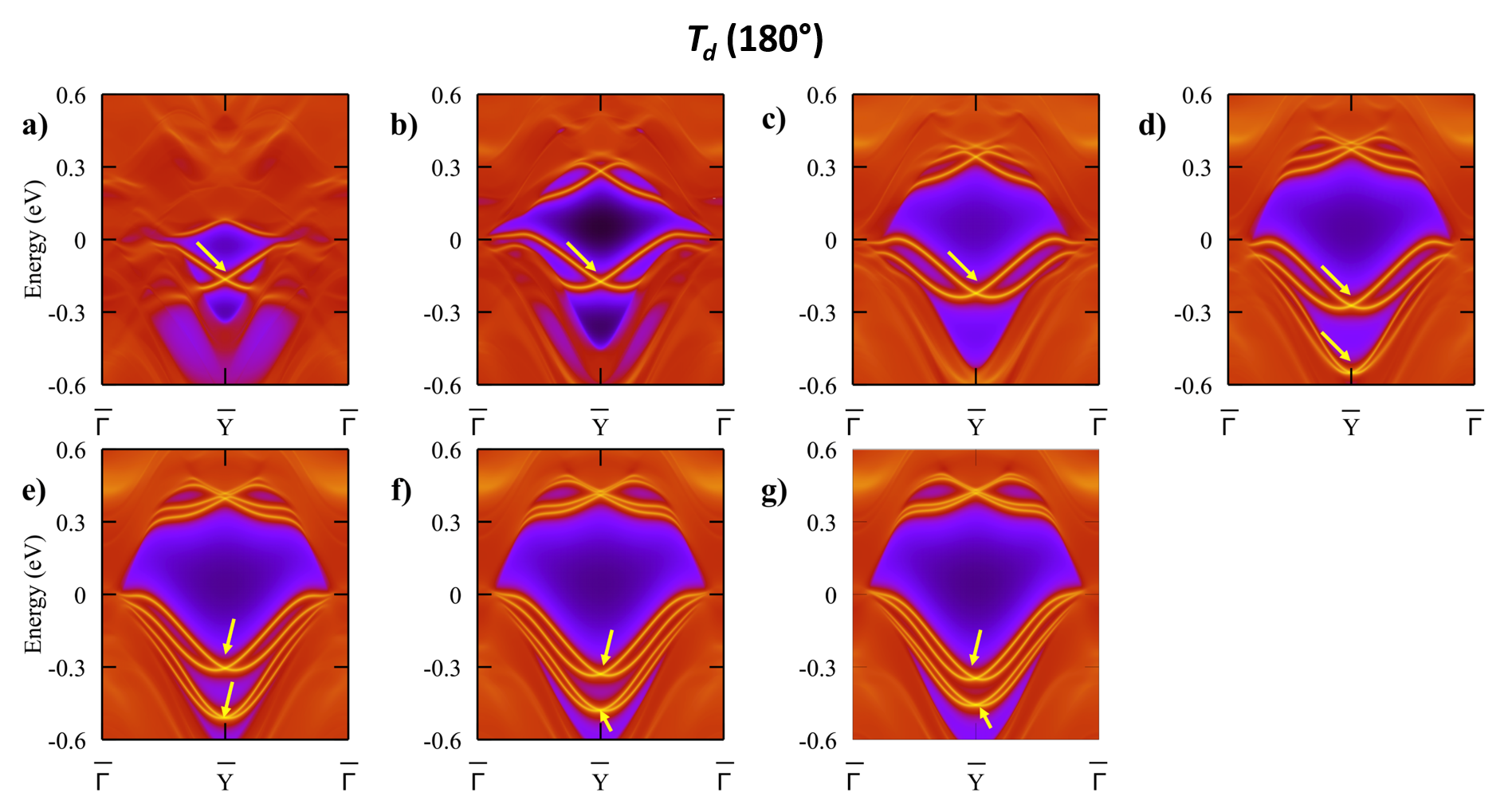}
		\caption{\label{Figs_DFT_4} {Edge states along  (010) direction in as-grown bilayer T$_d$ phase at different interlayer distance.} {(a)} d = 6.5 \AA,  {(b)} d = 6.75 \AA, {(c)} d = 7.0 \AA, {(d)} d = 7.25 \AA, {(e)} d = 7.5 \AA, {(f)} d = 7.75 \AA, {(g)} d = 8.0 \AA.}
	\end{figure*} 
	\begin{figure*}
		\centering
		\includegraphics[width=16cm]{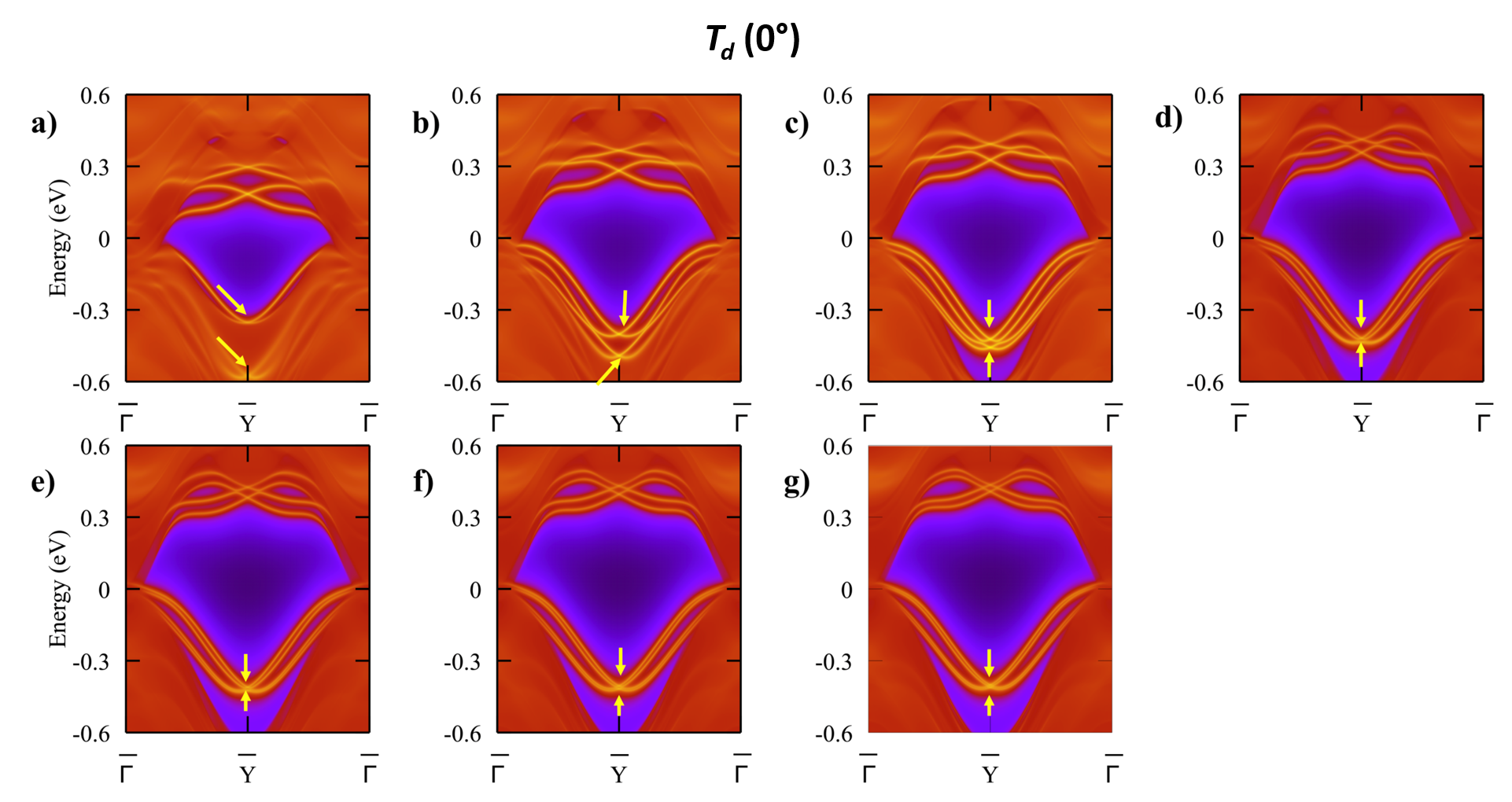}
		\caption{\label{Figs_DFT_5} {Edge states along  (010) direction in tbL ($\phi=0^{\circ}$) in T$_d$ phase at different interlayer distance.} {(a)} d = 6.5 \AA,  {(b)} d = 6.75 \AA, {(c)} d = 7.0 \AA, {(d)} d = 7.25 \AA, {(e)} d = 7.5 \AA, {(f)} d = 7.75 \AA, {(g)} d = 8.0 \AA.}
	\end{figure*} 
	\begin{figure*}
		\centering
		\includegraphics[width=16cm]{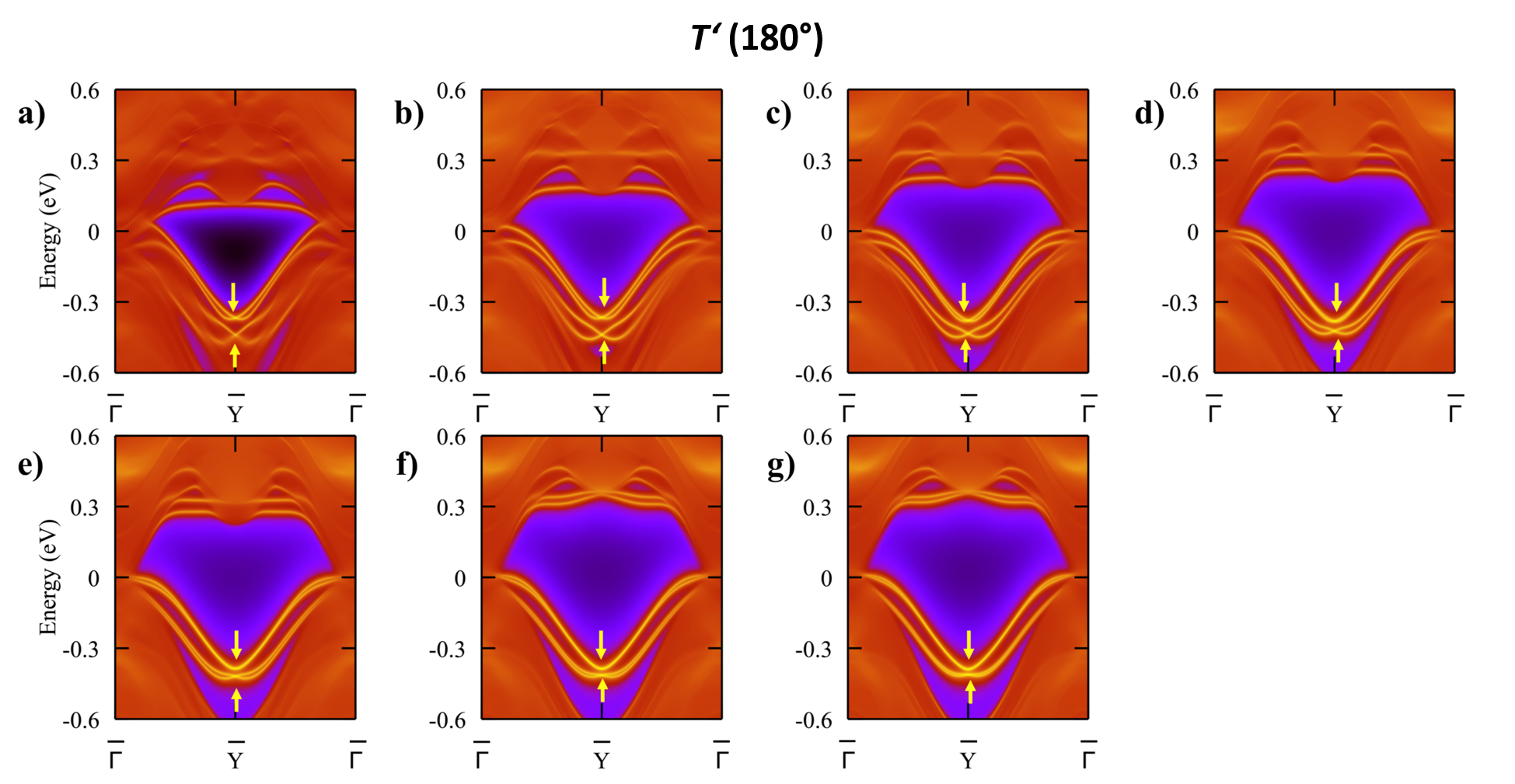}
		\caption{\label{Figs_DFT_6} {Edge states along  (010) direction in as-grown bilayer T' phase at different interlayer distance.} {(a)} d = 6.5 \AA,  {(b)} d = 6.75 \AA, {(c)} d = 7.0 \AA, {(d)} d = 7.25 \AA, {(e)} d = 7.5 \AA, {(f)} d = 7.75 \AA, {(g)} d = 8.0 \AA.}
	\end{figure*} 
	
	\begin{figure*}
		\centering
		\includegraphics[width=16cm]{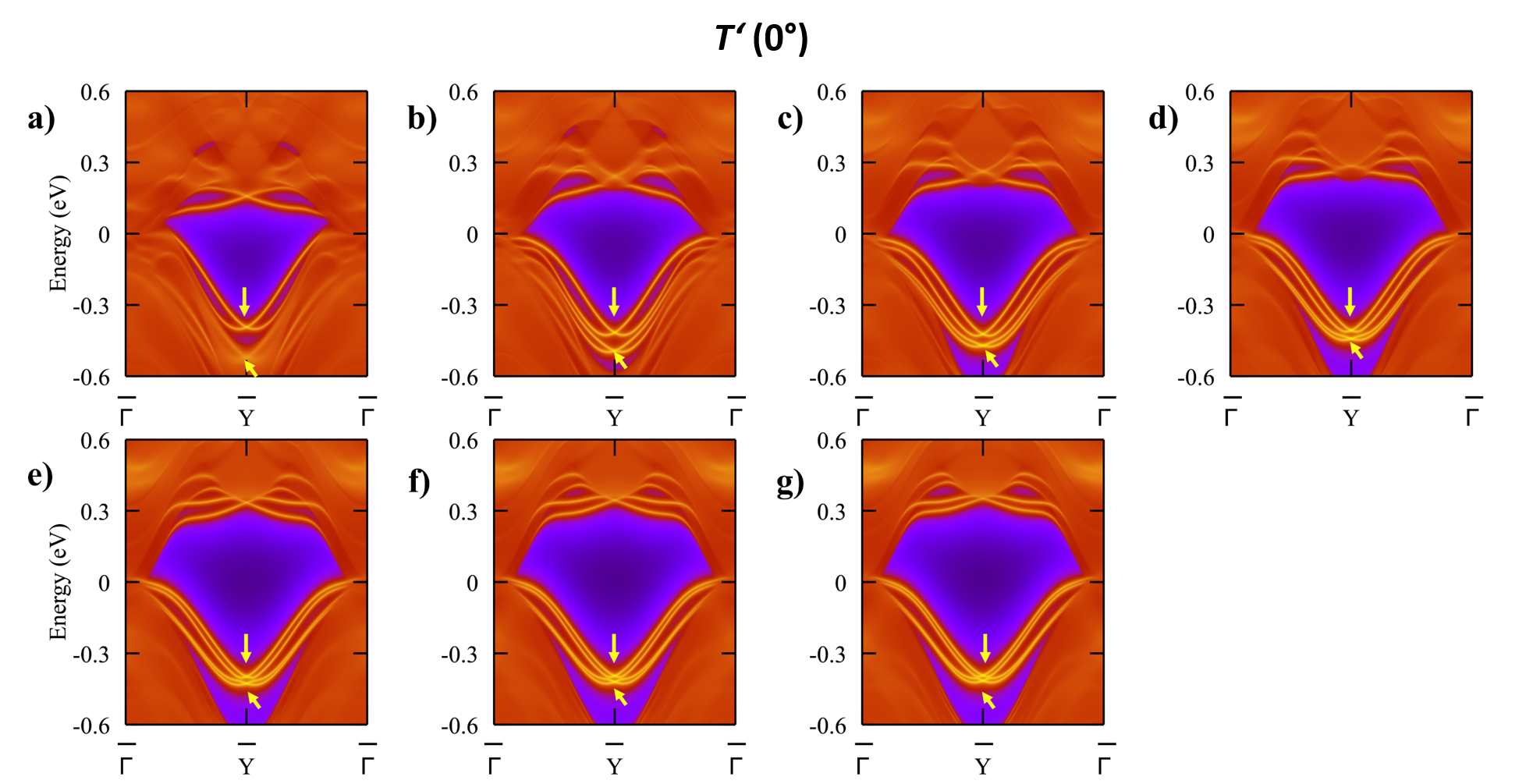}
		\caption{\label{Figs_DFT_7} {Edge states along  (010) direction in $\phi=0^{\circ}$ in T' phase at different interlayer distance.} {(a)} d = 6.5 \AA,  {(b)} d = 6.75 \AA, {(c)} d = 7.0 \AA, {(d)} d = 7.25 \AA, {(e)} d = 7.5 \AA, {(f)} d = 7.75 \AA, {(g)} d = 8.0 \AA.}
	\end{figure*} 
	\begin{figure*}
		\centering
		\includegraphics[width=16cm]{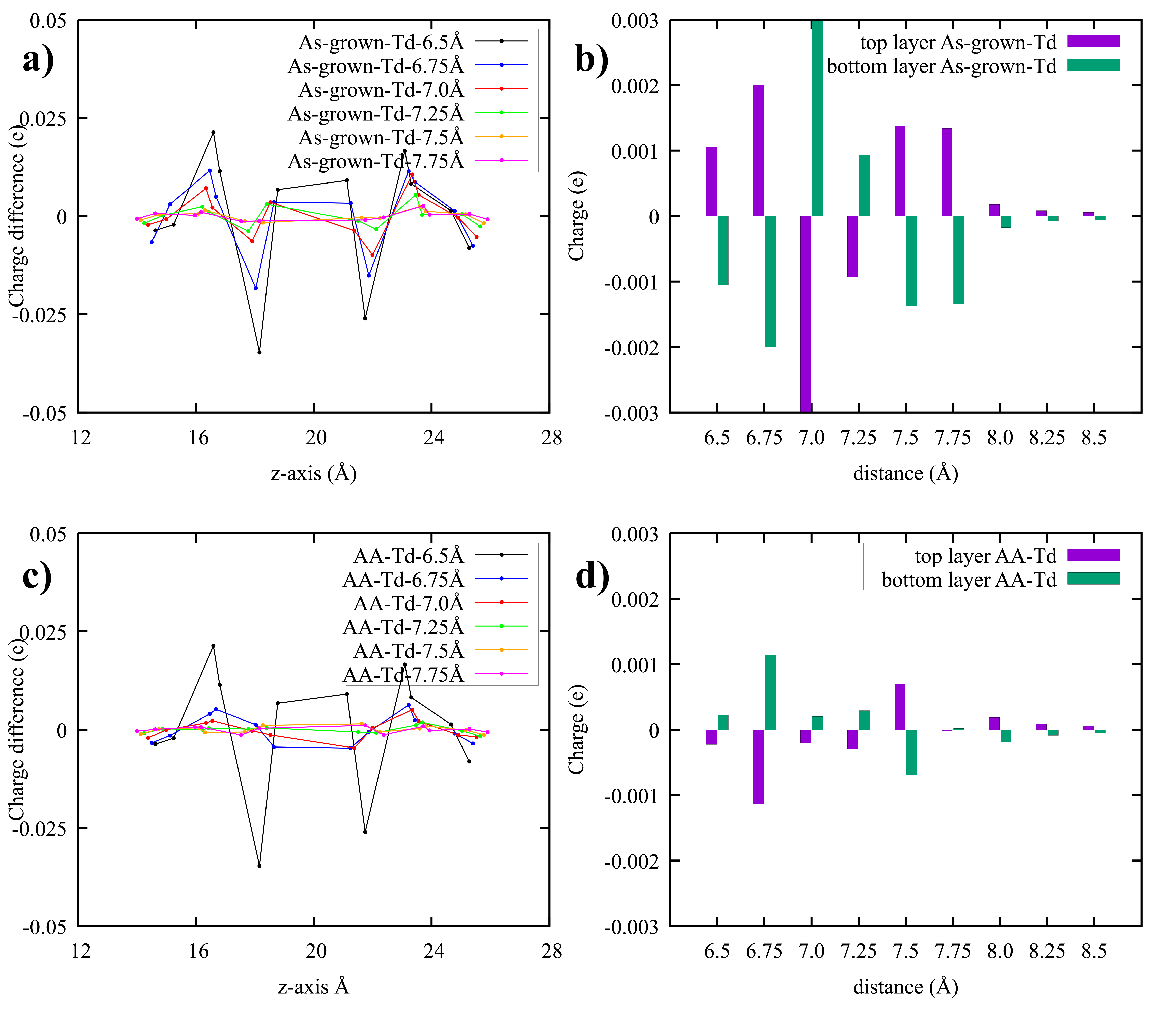}
		\caption{\label{Figs_DFT_8} {Charge transfer in T$_d$ bilayers.} {(a)} as-grown bilayer charge difference, {(b)} amount of charge transfer of the individual layers in as-grown bilayer. {(c)} $\phi=0^{\circ}$ bilayer charge difference, {(d)} amount of charge transfer of the individual layers for $\phi=0^{\circ}$ bilayer. The charge transfer is calculated as $\rho$ = $\rho_{\rm total}$ - $\rho_{\rm top}$ -  $\rho_{\rm bottom}$. The charge being transferred is calculated as the sum of the charge different for the top layer and bottom layer.}
	\end{figure*} 
	\begin{figure*}
		\centering
		\includegraphics[width=12cm]{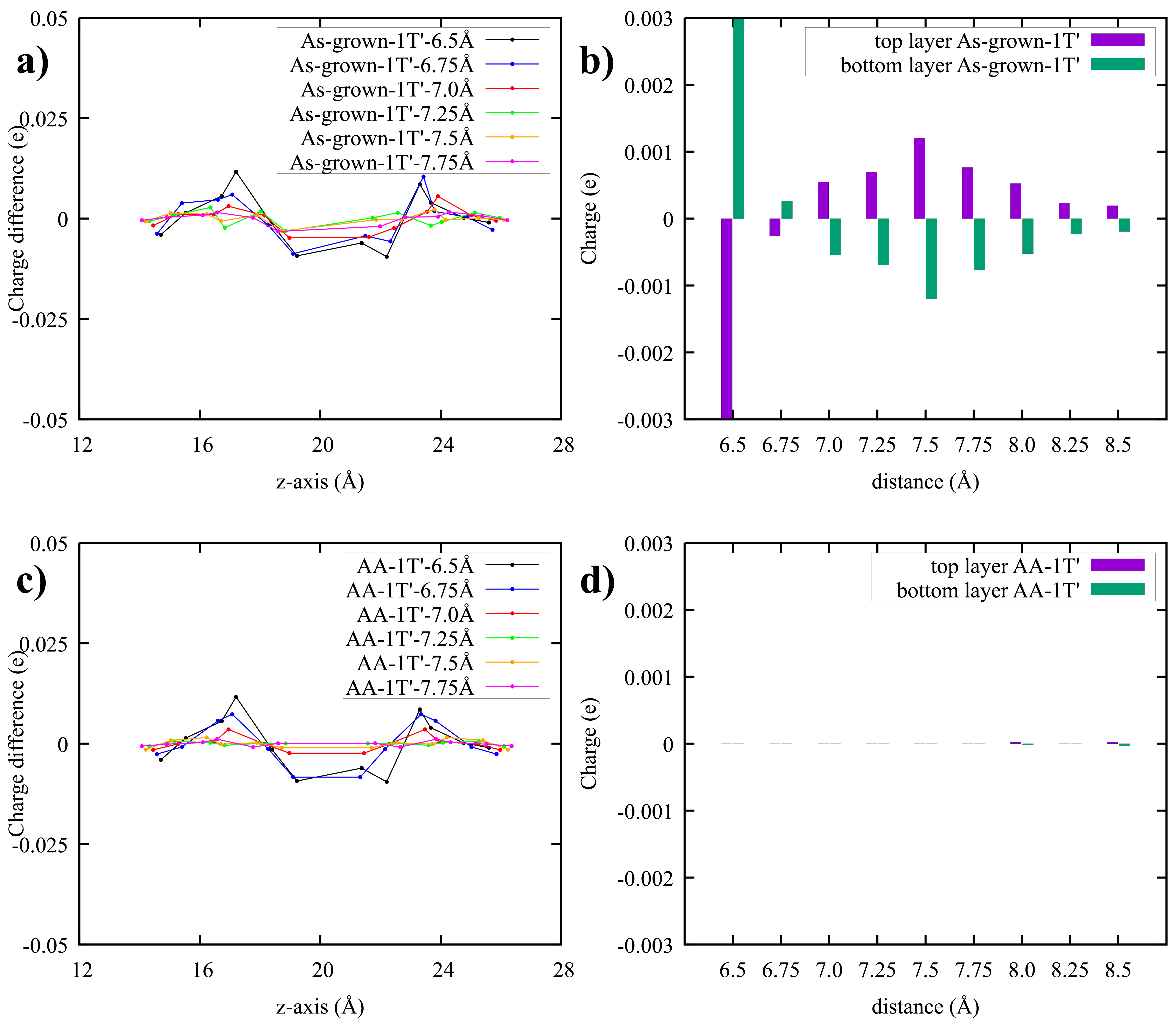}
		\caption{\label{Figs_DFT_9} {Charge transfer in T' bilayer.} {(a)} as-grown bilayer charge difference, {(b)} amount of charge transfer of the individual layers in as-grown bilayer. {(c)} $\phi=0^{\circ}$ bilayer charge difference, {(d)} vanishing
			charge transfer in $\phi=0^{\circ}$ bilayer due to the preserved symmetry.}
	\end{figure*} 
	
	\begin{table}[!ht]
		\begin{tabular}{c c} 
			\hline\hline
			Structure & Spacegroup  \\
			\hline
			Monolayer T' &  P2$_1$/m (11)\\
			AA-stacked bilayer T' & P2$_1$/m (11) \\
			As-grown bilayer T' & Pm (6) \\\hline 
			
			Monolayer T$_d$ & Pm (6)\\
			AA-stacked bilayer T$_d$ & Pm (6) \\
			As-grown bilayer T$_d$ & Pm (6) \\
			\hline
		\end{tabular}
		\caption{\label{table_structure} Crystal structures and corresponding space groups (space group numbers in parentheses) of the studied \WTe configurations in T$_d$ and T' phases}
	\end{table}

	{DFT calculations were performed using the VASP PAW potentials \cite{SI-vasp_paw1,SI-vasp_paw2}, and van der Waals (vdW) functional opt86b \cite{SI-Lee10,SI-Klimes11}, an energy cut off of {260 eV} and a dense $\Gamma$ centered 12$\times$10$\times$1 \textit{\textbf{k}}-mesh. 
		For our structures, we consider the monolayer and bilayer configurations in the T' (Fig. \ref{Figs_DFT_2}) and T$_d$ (Fig. \ref{Figs_DFT_1}) phases. 
		In all calculations, the lattice parameters for the monolayer and bilayer structures were set at $a = 3.421\,$\AA\ and $b = 6.236\,$\AA. 
		In our notation, the T$_d$ phase corresponds to an orthorhombic spacegroup Pmn2$_1$ (space group 31) with $\alpha=\beta=90^{\circ}$ in the bulk, and the T' phase belongs to the monoclinic spacegroup P2$_1$/m (space group 11) with a small tilt angle $\beta=93.917^{\circ}$ and $\alpha=90^{\circ}$, similar to Ref. \onlinecite{SI-Xu18}.
		To meet our definition of the {\sl a} lattice vector in the main text to be shorter than the {\sl b} lattice vetor, we rotate the structure by 90$^{\circ}$, resulting in $\alpha=86.083$$^{\circ}$ and $\beta=90$$^{\circ}$ in the T' phase. 
		A vacuum gap of $40\,$\AA\ was applied to all the configurations. 
		For bilayer configurations, we consider the as-grown bilayer as well as $0^{\circ}$ bilayer
		for both, T$_d$ and T'  structures.
		A summary of the structural symmetries for the monolayer and bilayer with different stacking is shown in table \ref{table_structure}. In addition, for the different stacking in the T$_d$ and T' phase, we vary the interlayer distance in steps of $0.5\,$\AA\ to study its effect on interlayer interaction, charge transfer and topology. The interlayer distance is defined as the distance between the W atoms in top and bottom layers. 
		Effects of relaxation were not found to change the topology, which is preserved by the overall symmetry. 
		As such, we only report results for the case where internal coordinates are unrelaxed. 
		The topology and edge states were calculated using the WannierTools program \cite{SI-Wu_theory_2018} based on projecting a tight-binding Hamiltonian for the low-energy bands around the Fermi-level onto W {\sl d} orbitals and Te {\sl p} orbitals using the Wannier90 package \cite{SI-Mostofi2008}. The charge transfer effect was investigated using the Bader analysis \cite{SI-Tang2009} with the charge difference $\rho$ = $\rho_{\rm total}$ - $\rho_{\rm top}$ -  $\rho_{\rm bottom}$.} 
	
	{The effect of charge transfer is calculated to be strongest in the as-grown T$_d$ bilayer. The charge transfer in tBL T$_d$ is smaller than in the corresponding T' tBL bilayer. In AA-stacked T' tBL, the charge transfer effect is completely absent. In the as-grown T' bilayers, there is a strong charge transfer between the layers, due to the lack of inversion symmetry, resulting in a polar metal. This is in stark contrast to the AA-stacking where the inversion symmetry is preserved, indicating weak coupling between the layers. In the T$_d$ phase, a charge transfer is present in both of the AA and as-grown bilayer due to their common lack of inversion symmetry, which also indicates the existence of the polar metal state in these configuration. However, in the as-grown T$_d$ bilayer a much stronger charge transfer effect is observed, indicating strong coupling between the layers. In contrast, the $\phi=0^{\circ}$ stacking results in a significantly smaller charge transfer, thus suggesting a weaker interlayer coupling.} 
	
	{Based on our calculation, ML \WTe in the T$_d$ and T' phase are topologically non-trivial with $\mathbb{Z}_2$ = 1 even though they belong to different symmetry classes (table \ref{table_structure}). As a result, they both exhibit a topological edge state when cutting along the $b$-axis (Fig. \ref{Figs_DFT_3}). Based on our DFT calculations, the edge states for, both, monolayer T$_d$ and T' are mostly located in the valence bands, similar to previous theoretical study of the edge state in 1T' monolayers \cite{SI-Lau19}. In addition, the prediction of the non-trivial  $\mathbb{Z}_2=1$ in ML \WTe in the T$_d$ phase is also consistent with recent model Hamiltonian prediction \cite{SI-Garcia20}.}
	
	{For the bilayer configurations in the T$_d$ and T' phases, we find the stacking geometry and interlayer distance to be important factors to determine the topology and charge transfer between the layers. In the T$_d$ phase, the topology  of the as-grown bilayer undergoes a transition from a $\mathbb{Z}_2$ = 1 to $\mathbb{Z}_2$ = 0 when we change the layer distances from being far apart to being extremely close together (Fig. \ref{Figs_DFT_4}). In contrast, in the AA-stacked T$_d$ bilayer $\mathbb{Z}_2=0$ globally regardless of the interlayer distance. Nevertheless, the electronic structure of the edge states varies significantly as function of interlayer distance. As shown in Fig. \ref{Figs_DFT_4}, the edge state in the as-grown T$_d$ bilayer forms a single  band when the layers are close, which transition to two edge bands at a critical interlayer distance of $7.0\,$\AA. This is in agreement with the observed topological transition. The position of the two edge states remains largely unchanged above the critical interlayer distance of $7.0\,$\AA. In the AA-stacked T$_d$ bilayers, the global $\mathbb{Z}_2$ invariant is equal to zero, but the two edges show a strong hybridization as the layers move further apart \ref{Figs_DFT_5}.
		In addition, because of the inversion symmetry breaking in both stacking configurations in the T$_d$ phase (table \ref{table_structure}), the charge transfer effect is predicted to exist in the as-grown and AA-stacking, however the strength of this effect is shown to be reduced in the AA-stacking (Fig. \ref{Figs_DFT_8}).}
	
	{
		In the T' phase, the inversion symmetry is preserved in the AA-stacking, thus a charge transfer only occurs in the as-grown bilayer (Fig. \ref{Figs_DFT_9}).
		In addition, even though there is no topological transition happening in the as-grown T' bilayer as function of interlayer distance, the edge states also exhibit similar behavior as the T$_d$ phase when the two layers are further apart. Specifically, in the AA-stacking, the two layers' edge states hybridize (Fig. \ref{Figs_DFT_7}) as the two layers are further apart while the edge states in the as-grown bilayer are largely unaffected with respect to the different interlayer distance.}
	
	{In our experiments, even though we can not distinguish between the T$_d$ and T' phases for the different bilayers, our DFT calculations suggest that only the stacking geometry and the interlayer distance are important in determining the topology and the charge transfer. Specifically, the AA-stacking in both of the T$_d$ and T' phase shows a significantly reduced charge transfer effect in comparison to the as-grown bilayers, thus suggesting a less metallic behavior in the AA-stacking bilayers compared to the as-grown structures regardless of the crystal symmetries. This DFT prediction is consistent with the experimental observation of an overall smaller {\sl dI/dV} signal in the twisted bilayer compared to the as-grown bilayer (compare Fig. 2 and 4 of the main text). In addition, the charge transfer for the AA-stacking is shown to be diminished significantly in our DFT calculation in both of the T$_d$ and T' phases. Experimentally, we observe the interlayer distance in the twisted BL to be larger than for the as-grown BL which supports the absence of a significant charge transfer in the measured tBL.}
	
	{In addition to the reduced charge transfer in the AA-stacking, our DFT calculations also predict a strong hybridization between the edge states on the individual layers when the interlayer separation distance is large in the AA-configuration for both T$_d$ and T' phases. As a result, since the interlayer distance is large for the tBL as measured in our experiments, the QSH edge-states can appear as the sum over the two layers as shown in Fig. \ref{Figs_DFT_5} and Fig. \ref{Figs_DFT_7} . Specifically, the overall topology of the tBL can be considered as a system of two stacked strong $\mathbb{Z}_2$ topological insulators due to the weak interlayer coupling, resulting from a large interlayer distance and twist angle rotation. Consequently, such a system can posses a unique four-fold Dirac point due to the overlapping of the individual Dirac points located on the two different layers (Fig. \ref{Figs_DFT_5} and Fig. \ref{Figs_DFT_7}). As a result, in scanning tunneling experiments of tBL's (with small rotational misalignement) the edge states are expected to be similar to that of the ML in agreement with Figs. 2e and f in the main text.}

	\bibliographystyle{naturemag}

\end{document}